\documentclass[twocolumn, journal]{IEEEtran}

\IEEEoverridecommandlockouts

\usepackage{color}
\usepackage{graphicx}
\usepackage{amsmath}
\usepackage{amssymb}
\usepackage{algorithm}
\usepackage{algorithmic}
\usepackage{amsmath}
\usepackage{multirow}
\usepackage{booktabs}
\usepackage{array}
\usepackage{amsthm}
\usepackage{stfloats}
\usepackage{caption}
\usepackage{bm}
\usepackage{lineno}

\newcommand{\be}{\begin{equation}}
\newcommand{\ee}{\end{equation}}
\newcommand{\bea}{\begin{eqnarray}}
\newcommand{\eea}{\end{eqnarray}}
\newcommand{\ba}{\begin{array}}
\newcommand{\ea}{\end{array}}

\newcommand{\non}{\nonumber}

\makeatletter

\newcommand{\Rmnum}[1]{\expandafter\@slowromancap\romannumeral #1@}
\makeatother

\newcommand{\RNum}[1]{\uppercase\expandafter{\romannumeral #1\relax}}

\title{IRS-assisted Multi-cell Multi-band Systems: Practical Reflection Model and Joint Beamforming Design
\thanks{W. Cai, R. Liu, M. Li, and Y. Liu are with the School of Information and Communication Engineering, Dalian University of Technology, Dalian 116024, China (e-mail: wenhaocai@mail.dlut.edu.cn; liurang@mail.dlut.edu.cn; mli@dlut.edu.cn; yangliu\_613@dlut.edu.cn).}
\thanks{Q. Wu is with the State Key Laboratory of Internet of Things for Smart City, University of Macau, Macau 999078, China (email: qingqingwu@um.edu.mo).}
\thanks{Q. Liu is with the School of Computer Science and Technology, Dalian University of Technology, Dalian 116024, China (e-mail: qianliu@dlut.edu.cn).}}
\author{Wenhao Cai,
        Rang Liu,~\IEEEmembership{Graduate Student Member,~IEEE,}
        Ming Li,~\IEEEmembership{Senior Member,~IEEE,}\\
        Yang Liu,~\IEEEmembership{Member,~IEEE,}
        Qingqing Wu,~\IEEEmembership{Senior Member,~IEEE,}
        and Qian Liu,~\IEEEmembership{Member,~IEEE}}

\begin{document}
\maketitle
\thispagestyle{empty}
\begin{abstract}
Intelligent reflecting surface (IRS) has been regarded as a promising and revolutionary technology for future wireless communication systems owing to its capability of tailoring signal propagation environment in an energy/spectrum/hardware-efficient manner.
However, most existing studies on IRS optimizations are based on a simple and ideal reflection model that is impractical in hardware implementation, which thus leads to severe performance loss in realistic wideband/multi-band systems.
To deal with this problem, in this paper we first propose a more practical and more tractable IRS reflection model that describes the difference of reflection responses for signals at different frequencies.
Then, we investigate the joint transmit beamforming and IRS reflection beamforming design for an IRS-assisted multi-cell multi-band system.
Both power minimization and sum-rate maximization problems are solved by exploiting popular second-order cone programming (SOCP), Riemannian manifold, minimization-majorization (MM), weighted minimum mean square error (WMMSE), and block coordinate descent (BCD) methods.
Simulation results illustrate the significant performance improvement of our proposed joint transmit beamforming and reflection design algorithms based on the practical reflection model in terms of power saving and rate enhancement.
\end{abstract}


\maketitle
\vspace{-0.4 cm}
\section{Introduction}
With the rapid development of wireless communication networks and the popularizing of various intelligent devices, the demands for high transmission rate and low latency have also been exponentially growing in the last decades \cite{6G}.
To accommodate these constantly increasing demands, several key technologies, such as heterogeneous dense networks, massive multiple-input multiple-output (MIMO), and millimeter-wave (mmWave) communications, have been proposed to improve system performance for the fifth-generation (5G) and beyond communication networks \cite{5G1, 5G2}.
However, it seems that we are approaching the limit of the theories obtained from such technologies.
Furthermore, the required high hardware complexity and consequently high energy consumption in practical implementations remain the bottlenecks for large-scale deployment \cite{Tutorial}.
Therefore, new technologies are desired to provide fundamental advances for future wireless networks in a more energy/spectrum/hardware-efficient fashion.

The innovative concept of intelligent reflecting surface (IRS) has recently emerged as such a promising technology \cite{Tutorial}-\cite{RIS6}.
An IRS is a two-dimensional (2D) planar array consisting of numerous reflecting elements, which are implemented by reconfigurable electromagnetic (EM) internals with very low power consumption.
Each reflecting element independently adjusts the phase-shift and amplitude of incident EM waves in a programmable manner, which collaboratively achieves reflection beamforming and reshapes propagation environments for wireless communications.
Deploying an IRS in existing communication networks brings the capability of creating a favorable propagation environment and provides more degrees of freedom (DoFs) for network optimizations.
Furthermore, these lightweight, hardware-efficient, and cost-effective reflecting elements provide the IRS with portability and mobility for various practical applications and enable large-scale IRSs to produce higher passive beamforming gains to significantly improve the transmission quality of service (QoS).
Therefore, IRS has been envisioned as a revolutionary technology owing to its capability of creating a smart and reconfigurable wireless propagation environment in a hardware-efficient way and has drawn significant attentions within both industry and academic communities \cite{Tutorial}.

In order to take advantage of the IRS mentioned above, extensive researches on deploying IRS in various wireless communication systems have been conducted to improve the performance under different metrics.
By judiciously adjusting IRS elements, the reflected signals are intelligently elaborated to achieve performance improvement in terms of spectral efficiency \cite{IRS_spectral}, energy efficiency \cite{IRS_energy efficiency}, transmit power \cite{Two-stage, IRS_Liuxing}, sum-rate \cite{IRS_MIMO}-\cite{Hongyu}, etc., for single-user/multi-user MIMO/multi-input single-output (MISO) \cite{IRS_spectral}-\cite{IRS_weighted sum-rate}, wideband orthogonal frequency division multiplexing (OFDM) \cite{Hongyu}, or multi-cell systems \cite{IRS_multi-cell}.
In addition, researchers have explored the designs of IRS with low-resolution phase-shift \cite{IRS_discrete}  and phase error \cite{IRS_phase error} or under imperfect channel state information (CSI) \cite{IRS_UCsi}.
Moreover, IRS has been applied in various novel applications, such as physical layer security \cite{IRS_Secure, IRS_Secure2}, index modulation \cite{IRS_index modulation}, passive information transmission \cite{IRS_passive information transmission}, etc.

In the above applications, it is assumed that each reflecting element has an ideal IRS reflection model, which induces the same constant amplitude yet variable phase-shift response to the incident signals.
Based on this ideal IRS reflection model, a lot of existing algorithms, e.g., majorization minimization (MM), Riemannian manifold optimization, semidefinite relaxation (SDR), etc., have been readily employed in IRS reflection optimizations.
Unfortunately, such an ideal IRS is difficult to be realized by current hardware circuit techniques.
Therefore, existing designs with the ideal reflection model inevitably suffer from severe performance loss in realistic systems since the responses of practical hardware circuits are quite different from the ideal one \cite{Abeywickrama TCOM 2020}-\cite{IRS_HardwareE}.

In order to unlock the full potential IRS in realistic systems, many different practical IRS reflection models are derived to illustrate the mechanism that affects the reflection coefficient \cite{Abeywickrama TCOM 2020}-\cite{Practical_phase2}, \cite{irs_practical}.
The seminal work \cite{Abeywickrama TCOM 2020} presented a two-dimensional amplitude-phase reflection model, which shows a fundamental relationship between the reflection amplitude and phase-shift for narrowband systems.
However, it has been verified that the reflection amplitude and phase-shift vary with the frequency of incident signals \cite{IRS_Hardware}.
This two-dimensional reflection model, which does not consider the effect of signal frequency, cannot be readily applied in wideband/multi-band systems.
To tackle this issue, in our previous work \cite{Practical_phase}, we analyzed the responses to signals at different frequencies and established a three-dimensional amplitude-frequency-phase reflection model.
Since this sophisticated model brings huge difficulties for joint beamforming and reflection designs, a simplified version was developed in \cite{Practical_phase2} for typical wideband OFDM systems.
Besides, the authors in \cite{irs_practical} derived a more general reflection model using rigorous scattering parameter network analysis, which mainly focuses on simple IRS-aided communication systems and thus leads to a complicated model for a certain complicated communication system, e.g., the multi-cell multi-band system considered in this paper.
Therefore, the existing IRS reflection model is too complicated to be used for the joint beamforming and IRS reflection designs. Deriving a more straightforward and more tractable practical IRS reflection model for the multi-cell multi-band systems remains an open problem.

Motivated by the above, in this paper we propose a simplified practical IRS reflection model for IRS-assisted multi-cell multi-band systems and then investigate associated joint transmit beamforming and IRS reflection beamforming designs.
Specifically, multiple base stations (BSs), which belong to the same service provider but operate at different frequency bands, simultaneously serve the users in their cells with the aid of one IRS.
To the best of our knowledge, this problem has not been investigated in the literature yet.
Our main contributions are summarized as follows.
\begin{itemize}
  \item We first analyze the IRS phase-shift response to the incident signals at different frequencies and provide a lean practical frequency-dependent IRS reflection model.
      Compared with our previous work, this newly proposed model significantly reduces the required computational complexity and is more suitable for realistic multi-cell multi-band systems.
  \item Then, we investigate the power minimization problem, which aims to minimize the total transmit power subject to the signal-to-interference-plus-noise ratio (SINR) constraints of all users and the practical IRS reflection model.
      A three-step algorithm is proposed to solve for the transmit beamforming and the practical IRS reflection beamforming by utilizing the second-order cone programming (SOCP), the Riemannian manifold optimization, and the minimization-majorization (MM) method after some sophisticated transformations.
  \item The sum-rate maximization problem is also investigated to maximize the sum-rate subject to the total transmit power constraint and the practical IRS reflection model.
      In order to effectively handle this non-convex problem, we exploit the weighted minimum mean square error (WMMSE) approach to convert the original problem into a solvable multi-variable optimization, which is then handled by the block coordinate descent (BCD) method.
      Based on the proposed lean practical frequency-phase model, the closed-form solutions of IRS phase-shift and service selection of each element can be simultaneously obtained for the sum-rate maximization problem, which can significantly improve the efficiency of the algorithm.
  \item Finally, extensive simulation results are illustrated to show the effectiveness of our proposed algorithms and validate the significant performance improvement achieved by considering the proposed practical IRS reflection model in the designs for multi-cell multi-band communication networks.
\end{itemize}

\begin{figure}[t]
\centering
  \includegraphics[height = 2.3 in]
  {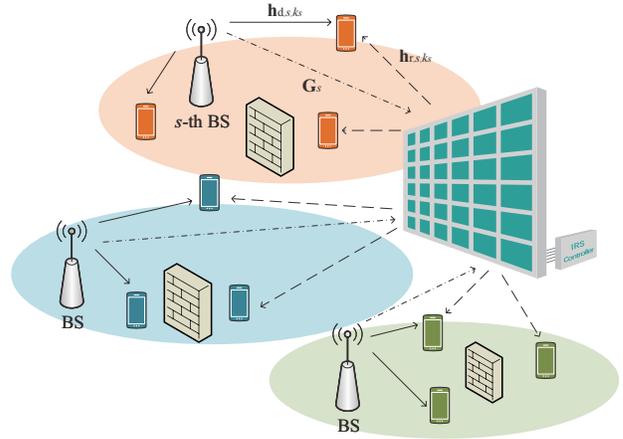}
  \vspace{0.2 cm}
  \caption{An IRS-assisted multi-cell multi-band system.}\label{fig:System model}
\end{figure}

\textit{Notation}:
Boldface lower-case and upper-case letters indicate column vectors and matrices, respectively. $\mathbb{C}$ and $\mathbb{R}^+$ denotes the set of complex and positive real numbers, respectively. $(\cdot)^*$, $(\cdot)^T$, $(\cdot)^H$, and $(\cdot)^{-1}$ denote the conjugate, transpose, conjugate-transpose operations, and the inversion of a matrix, respectively. $\mathbb{E}\{\cdot\}$ and $\mathfrak{R}\{\cdot\}$ denote statistical expectation and the real part of a complex number, respectively. $\mathbf{I}_L$ indicates an $L \times L$ identity matrix. $||\mathbf{a}||$ and $||\mathbf{a}||_0$ denote the $\ell_2$ norm and $\ell_0$ norm of a vector $\mathbf{a}$, respectively. $\odot$ denotes the Hadamard product. In addition, $\angle{\theta}$ denotes the angle of a complex number $\theta$.

\section{System Model and Practical Reflection Model }
\subsection{System Model}
We consider an IRS-assisted multi-cell multi-band wireless communication system as shown in Fig. \ref{fig:System model}, where an IRS composed of $M$ reflecting elements is deployed to simultaneously assist the downlink communications in $S$ cells.
Specifically, in the $s$-th cell, $s = 1, 2, \ldots, S$, the BS equipped with $N_\text{t}$ transmit antennas serves $K_s$ single-antenna users at the frequency $f_s$.
Let $\mathcal{S} \triangleq \{1, \ldots, S\}$ denote the set of BSs, $\mathcal{K}_s \triangleq \{1, \ldots, K_s\}$ denote the set of the users served by the $s$-th BS, and $\mathcal{M} \triangleq \{1, \ldots, M\}$ denote the set of IRS reflecting elements.
Without loss of generality, we assume $f_1 > f_2 > \cdots > f_S$ in the considered multi-cell multi-band system.
In addition, the IRS is controlled by an IRS controller through a dedicated control link and only the first-order reflection is considered due to significant path loss.

Denote $\mathbf{z}_{s} \triangleq [z_{s,1}, \ldots, z_{s,K_s}]^T \in \mathbb{C}^{K_s}$ as the transmitted symbols for the users served by the $s$-th BS, $\mathbb{E}\{\mathbf{z}_{s} \mathbf{z}_{s}^H\} = \mathbf{I}_{K_s}$, and $\mathbf{W}_s \triangleq [\mathbf{w}_{s,1}, \ldots, \mathbf{w}_{s,K_s}] \in \mathbb{C}^{N_\text{t}\times K_s}$ as the precoder matrix of the $s$-th BS, $\forall s \in \mathcal{S}, k_s\in\mathcal{K}_s$.
Since the BSs operate at different frequencies, the inter-cell interference can be easily eliminated through receiving filters at the users.
Thus, the received baseband signal at the $k_s$-th user served by the $s$-th BS can be expressed as
\begin{equation}
    y_{s,k_s} = \left( \mathbf{h}^H_{\text{r},s,k_s}\bm{\Theta}_s\mathbf{G}_s + \mathbf{h}^H_{\text{d},s,k_s} \right) \hspace{-0.2 cm} \sum_{k_s\in\mathcal{K}_s}\mathbf{w}_{s,k_s} z_{s,k_s} + n_{s,k_s},
    \label{eq:y}
\end{equation}
where $\mathbf{h}_{\text{r},s,k_s} \in \mathbb{C}^{M}$, $\mathbf{G}_{s} \in \mathbb{C}^{M \times N_\text{t}}$, and $\mathbf{h}_{\text{d},s,k_s} \in \mathbb{C}^{N_\text{t}}$ represent the baseband equivalent channels from the IRS to the $k_s$-th user, from the $s$-th BS to the IRS, and from the $s$-th BS to the $k_s$-th user, respectively.
It is noted that the quasi-static flat-fading Rayleigh channel model\footnote{The proposed practical IRS reflection model and associated design algorithms are also suitable for other channel fading models, e.g., Rician fading.} is adopted for all channels and we assume that all the CSIs are perfectly known at the BSs given existing efficient channel estimation approaches \cite{Channel1}-\cite{Channel9}. For example, the CSIs can be acquired at different BSs based on the received pilot signals from their users with acceptable training overhead.
$n_{s,k_s} \sim \mathcal{C}\mathcal{N}(0,\sigma^2)$ denotes the additive white Gaussian noise (AWGN) at the $k_s$-th user.
$\bm{\Theta}_s \triangleq \text{diag}\{\bm{\theta}_s\}$, and $\bm{\theta}_s \triangleq [\theta_{s,1}, \ldots, \theta_{s,M}]^T$ is the IRS reflection coefficient vector for the signals transmitted by the $s$-th BS, i.e., the signals at the frequency $f_s$.
We emphasize that with the same IRS settings, the IRS reflection vectors are different for the incident signals at different frequencies due to practical hardware characteristics, and they are inherently correlated as described in the next subsection.

With the received signal in (\ref{eq:y}), the SINR of the $k_s$-th user served by the $s$-th BS is given by
\begin{equation}
\gamma_{s,k_s} = \frac{\big|\big({\mathbf{h}^{H}_{\text{r},s,k_s}} \bm{\Theta}_s \mathbf{G}_{s} + \mathbf{h}_{\text{d},s,k_s}^H\big)\mathbf{w}_{s,k_s}\big|^2}{\sum^{j \in \mathcal{K}_s}_{j \neq k_s}\big|\big({\mathbf{h}^{H}_{\text{r},s,k_s}}\bm{\Theta}_s \mathbf{G}_{s} + \mathbf{h}_{\text{d},s,k_s}^H\big)\mathbf{w}_{s,j}\big|^2+\sigma^2}.
\end{equation}

\begin{figure}[t]
\centering
  \includegraphics[width= 3.1 in]{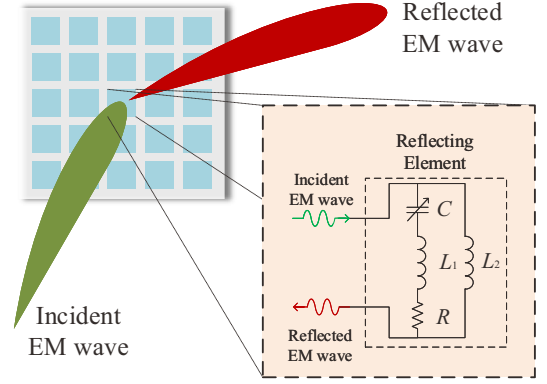}
  \caption{The equivalent circuit of a reflecting element.}\label{fig:electric_line}
  \vspace{-0.4 cm}
\end{figure}

\subsection{Practical IRS Reflection Model}
An IRS is typically implemented by a printed circuit board (PCB), in which semiconductor devices \cite{IRS_Hardware1}, e.g., positive-intrinsic-negative (PIN) diodes, are embedded to tune the reflection response by varying the impedance of each reflecting element.
As discussed in \cite{Abeywickrama TCOM 2020}-\cite{Practical_phase2}, the  reflection response of a reflecting element can be described by an equivalent parallel resonant circuit as shown in Fig. \ref{fig:electric_line}, where $L_1$, $L_2$, $C$, and $R$ denote the equivalent inductances, variable capacitance, and the loss resistance, respectively.
Thus, the impedance of a reflecting element can be written as
\begin{equation}
Z(C,f) = \frac{j2{\pi}f L_1(j2{\pi}f L_2+\frac{1}{j2{\pi}f C}+R)}{j2{\pi}fL_1+(j2{\pi}fL_2+\frac{1}{j2{\pi}f C}+R)}.\label{eq:gammaz}
\end{equation}
Then, the reflection coefficient $\theta$ that describes the effect of a reflecting element on the incident EM waves is given by
\begin{equation}
\theta(C, f) = \frac{Z(C,f)-Z_0}{Z(C,f)+Z_0},\label{eq:gamma}
\end{equation}
where $Z_0 = 377\Omega$ denotes the free space impedance.
From the microwave theory (\ref{eq:gamma}), we see that the reflection coefficient $\theta(C, f)$ is a function of the capacitance $C$ and the frequency $f$ of incident signals.
In other words, the reflection coefficient of each reflecting element is controlled by the variable capacitance and the same reflecting element exhibits different amplitude and phase-shift responses to the signals at different frequencies.
The creditability and correctness of the microwave theory can also be verified by the experimental and simulation results in \cite{IRS_Hardware}, \cite{IRS_HardwareE}.
We have analyzed this phenomenon and established a sophisticated three-dimensional amplitude-frequency-phase model in our initial work \cite{Practical_phase}, and further provided a simplified version for IRS optimization in wideband systems \cite{Practical_phase2}.
However, this practical IRS reflection model is composed of complex arc-tangent and \textit{Witch of Agnesi} functions, which are too complicated and inconvenient for the joint beamforming and IRS reflection designs in the considered multi-cell multi-band systems. This motivates us to derive a simpler approximated reflection model.

\begin{figure}[t]
\centering
  \includegraphics[width= 3.1 in]{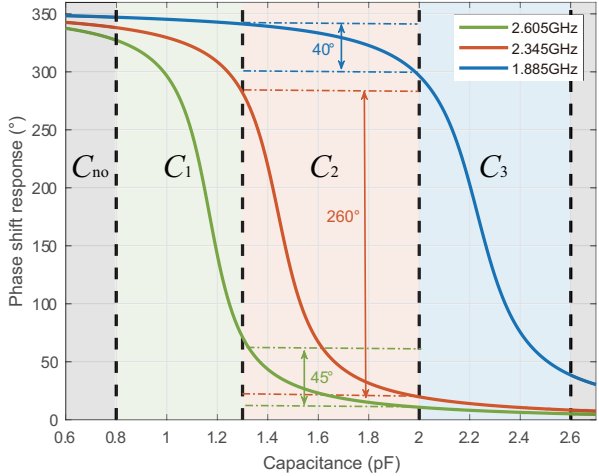}
  \caption{The phase-shift response versus capacitance for different frequencies ($S = 3$, $f_1 = 2.605$GHz, $f_2 = 2.345$GHz, $f_3 = 1.885$GHz).}
  \label{fig:Mode}
\vspace{-0.6 cm}
\end{figure}

To establish a simplified and tractable IRS reflection model, we first plot the phase-shift response $\angle \theta$ versus the capacitance in Fig. \ref{fig:Mode} according to (\ref{eq:gammaz}) and (\ref{eq:gamma}).
We take the scenario that $S = 3$, $f_1 = 2.605$GHz, $f_2 = 2.345$GHz, and $f_3 = 1.885 $GHz as an example\footnote{The proposed IRS reflection model can be easily extended to other frequency combinations and similar conclusions can be obtained.}, which adopts three available frequencies of China Mobile 4G network.
In addition, we choose a practical surface-mount diode SMV1231-079 \cite{IRS_Realize} with parameters $L_1 = 2.5$nH, $L_2 = 0.7$nH, and $R = 1\Omega$ to implement the IRS.
It can be noticed that the phase-shift response varies with the frequency of incident signal as analyzed above.
In addition, we observe that for a given frequency $f$, the phase-shift can be accurately tuned in a specific capacitance range but maintains almost the same in other ranges.
Furthermore, the tunable capacitance ranges of different frequencies are nearly not overlapped when the difference of frequencies is relatively large.
In other words, when we adjust the capacitance to provide a tunable phase-shift for a certain frequency, the phase-shift response for other frequencies is almost unchanged.
In this way, the capacitance values in Fig. \ref{fig:Mode} can be divided into four ranges (the green, orange, blue, and gray areas), which exhibit four different reflection responses for the signals of three different frequencies.

Take the orange area as an example.
When the capacitance varies from 1.3pF to 2pF, the phase-shift response for signals at 2.345GHz changes by $260^\circ$, while 1.885GHz changes by $40^\circ$ and 2.605GHz changes by $45^\circ$.
Therefore, we can approximately consider this capacitance range as that the reflecting element serves the BS operating at 2.345GHz by providing an ideal $(0, 2\pi]$ phase-shift and meanwhile exhibits a fixed $0/2\pi$ phase-shift for other BSs.
Since the IRS consists of lots of reflecting elements, the performance loss due to the approximation for each independent reflecting element is negligible.
Moreover, considering the hardware complexity and cost in practical implementations, the tunable phase-shift values are usually discrete (e.g., 1-bit or 2-bit resolution), which also supports our approximation.
Correspondingly, the green and blue areas indicate that the BSs operating at 2.605GHz and 1.885GHz are served by this reflecting element, respectively.
Meanwhile, in the gray areas, IRS provides fixed and uncontrollable phase-shifts to all the BSs. Therefore, IRS cannot generate effective beamforming for any BSs.

\begin{table}[!t]\normalsize
\centering\label{table:1}
\caption{A simplified representation of Fig. \ref{fig:Mode}.}
\vspace{-0.1 cm}
\begin{center}
\begin{tabular}{ c | c | c | c | c }
   \hline
      & $C_1$  & $C_2$ & $C_3$ & $C_\text{no}$\\
    \hline
    2.605 $\text{GHz}$ & $(0, 2\pi]$ & $0$         & $0$         & $0/2\pi$ \\
    \hline
    2.345 $\text{GHz}$ & $2\pi$      & $(0, 2\pi]$ & $0$         & $0/2\pi$\\
    \hline
    1.885 $\text{GHz}$ & $2\pi$      & $2\pi$      & $(0, 2\pi]$ &$0/2\pi$\\
    \hline
\end{tabular}\end{center}\vspace{-0.4 cm}
\end{table}

Based on the above descriptions, we simplify the phase-shift response for different operating frequencies in Table \RNum{1}, where $C_1$, $C_2$, $C_3$, and $C_\text{no}$ correspond to the capacitance ranges of the green, orange, blue, and gray areas in Fig. \ref{fig:Mode}, respectively.
We can clearly see that by controlling the capacitance of a reflecting element, a certain BS (i.e., the signal with a certain frequency) is selected to be served by this reflecting element with
an ideal (i.e., $0-2\pi$ tunable) phase-shift meanwhile the other BSs have a fixed $0/2\pi$ phase-shift, or none of the BSs are selected.
This frequency-selective characteristic of each reflecting element motivates us to describe a practical phase-shift response using the product of an ideal phase-shift and a binary indicator.
This simplified IRS phase-shift response transforms the extremely complicated nonlinear functional relationship between phase-shift and frequency into a binary functional relationship, which can further facilitate the joint transmit beamforming and IRS phase-shift design.
Thus, for the $m$-th reflecting element, we define its service selection vector as $\mathbf{a}_m \triangleq [a_{1,m}, \ldots, a_{S,m}]^T$, $a_{s,m} \in \{0,1\}$.
Specifically, $a_{s,m} = 1$ indicates that the $s$-th BS is selected to be served with a fully tunable phase-shift response provided by the $m$-th reflecting element, and $a_{s,m} = 0$ represents that the $m$-th reflecting element exhibits a fixed $0/2\pi$ phase-shift for the $s$-th BS.
For example, the four status in Table I can be represented by $\mathbf{a}_m = [1,0,0]^T$, $\mathbf{a}_m = [0,1,0]^T$, $\mathbf{a}_m = [0,0,1]^T$, and $\mathbf{a}_m = [0,0,0]^T$, respectively.
In addition, according to the above analysis, each reflecting element can serve at most one BS.
Thus, the constraint for the service selection vector of the $m$-th reflecting element can be expressed as
\begin{equation}
\left\| \mathbf{a}_m \right\|_0 \leq 1, \;\;a_{s,m} \in \{0,1\}, \;\;\forall s, m.
\end{equation}
We further define an $S \times M$ service selection matrix for the IRS as $\mathbf{A} \triangleq [\mathbf{a}_1, \ldots, \mathbf{a}_M]$.
It is noted that the $s$-th row of $\mathbf{A}$ is also the service selection  vector for the $s$-th BS and indicates which reflecting elements serve this BS.
For brevity, we denote the service selection vector for the $s$-th BS as $\widetilde{\mathbf{a}}_s \triangleq [a_{s,1}, \ldots, a_{s,M}]^T\in \{0,1\}^M$, which leads to
\begin{equation}
\mathbf{A} = [\mathbf{a}_1, \ldots, \mathbf{a}_M] = [\widetilde{\mathbf{a}}_1, \ldots, \widetilde{\mathbf{a}}_S]^T.
\end{equation}
Denote the ideal phase-shift vector $\bm{\phi}_s$ for the $s$-th BS as $\bm{\phi}_s \triangleq [\phi_{s,1}, \ldots, \phi_{s,M}]^T,\;\phi_{s,m} \in (0, 2\pi]$.
The practical phase-shift $\angle\bm{\theta}_s$ is thus given by
\begin{equation}
\angle\bm{\theta}_s = \bm{\phi}_s \odot \widetilde{\mathbf{a}}_s.
\end{equation}

Besides, as shown in \cite{Abeywickrama TCOM 2020}-\cite{Practical_phase2}, the reflection amplitude also varies with the frequency of signals.
However, for the signals at a certain frequency, i.e., transmitted by a certain BS, each reflecting element has the same reflection amplitude degradation, which can be compensated at the transmitter side using some advanced hardware techniques \cite{IRS_Hardware2}.
Therefore, neglecting the practical reflection amplitude response will not affect the optimizations for the considered multi-cell multi-band systems.
In this way, the practical reflection vector for the $s$-th BS can be written as $\bm{\theta}_s = \exp\left(\jmath\angle\bm{\theta}_s\right)$.
In summary, the practical IRS reflection model is formulated as
\begin{subequations}
\begin{align}
&\bm{\theta}_s = \exp\left(\jmath\bm{\phi}_s \odot \widetilde{\mathbf{a}}_s\right),\;\; \forall s, \\
&\phi_{s,m} \in (0, 2\pi], \;\;\forall s,\;m, \\
&\left\| \mathbf{a}_m \right\|_0 \leq 1, \;\;a_{s,m} \in \{0,1\}, \;\;\forall s,\;m.
\end{align}
\end{subequations}

Given the above system model and practical IRS reflection model, the joint transmit beamforming and IRS reflection designs for the power minimization and sum-rate maximization problems will be investigated in Sec. III and Sec. IV, respectively.

\section{Algorithm for Power Minimization Problem}
In this section, we aim to jointly optimize the transmit beamformers $\mathbf{W} \triangleq [\mathbf{W}_1,\ldots, \mathbf{W}_S]$ for all BSs, the ideal IRS phase-shifts $\bm{\Phi} \triangleq [\bm{\phi}_1, \ldots, \bm{\phi}_S]$ for all BSs, and the service selection matrix $\mathbf{A}$ to minimize the total transmit power, subject to users' SINR requirements and the practical reflection model.
Therefore, the power minimization problem is formulated as
\begin{subequations}
\label{eq:PM problem}
\begin{align}\label{eq:PM problem a}
   \min\limits_{\mathbf{W},\bm{\Phi},\mathbf{A}}~~&
   \sum_{s\in\mathcal{S}} \sum_{k_s\in\mathcal{K}_s}\left\|\mathbf{w}_{s,k_s}\right\|^2\\
   \label{eq:PM problem b}
   \textrm{s.t.}~~& \gamma_{s,k_s} \geq \Gamma_{s,k_s}, \;\;\forall s,\; k_s, \\
   \label{eq:PM problem c}
   &\bm{\theta}_s = \exp\left(\jmath\bm{\phi}_s \odot \widetilde{\mathbf{a}}_s\right),\;\; \forall s, \\
   \label{eq:PM problem d}
   &\phi_{s,m} \in (0, 2\pi], \;\;\forall s,\; m, \\
   \label{eq:PM problem e}
   &\left\| \mathbf{a}_m \right\|_0 \leq 1, \;\;a_{s,m} \in \{0,1\}, \;\;\forall s,\; m,
\end{align}
\end{subequations}
where $\Gamma_{s,k_s} > 0$ denotes the minimum SINR requirement of the $k_s$-th user served by the $s$-th BS.
The non-convex NP-hard problem (\ref{eq:PM problem}) is very difficult to solve due to the following three reasons.
First, the coupled variables brought by the constraints (9b) and (9c) make this multi-variable problem hard to solve.
Second, the frequency-selective model introduced by the constraint (9c) causes significant difficulty to the IRS phase-shift design.
Third, the practical reflection responses for different BSs are different and related by a binary service selection matrix in constraints (\ref{eq:PM problem c})-(\ref{eq:PM problem e}).
In order to tackle these difficulties, we utilize the characteristics of the proposed practical reflection model to decompose the original problem (\ref{eq:PM problem}) into two sub-problems and develop a three-step algorithm to solve the original problem as presented below.

As described in Sec. II-B, by tuning the capacitance, each reflecting element independently selects at most one BS to provide an ideal/tunable phase-shift.
This fact motivates us to decompose the original problem into service selection design and the joint transmit beamforming, ideal phase-shift design sub-problems.
In particular, in order to better evaluate the ability of each BS in exploiting IRS and determine the service selection, we first assume that the IRS is ideal/tunable for all BSs (i.e., $\widetilde{a}_s = \mathbf{1}, \forall s$) and jointly design the transmit beamforming and ideal phase-shifts for each BS.
Then, the service selection matrix $\mathbf{A}$ is optimized to select appropriate reflecting elements for serving each BS.
Finally, with the obtained $\mathbf{A}$ the ideal phase-shifts of selected reflecting elements and associated transmit beamforming are jointly designed for each BS.
It is noted that the joint transmit beamforming and ideal phase-shift design with ideal IRS reflection model (i.e., $\mathbf{A} = 1$) or the practical IRS reflection model follows the same route, which is presented in Sec. III-A in details.
In addition, the procedure of the service selection design is described in Sec. III-B.

\subsection{Joint Transmit Beamforming and Ideal Phase-Shift Design}
Based on the above analysis, with fixed $\mathbf{A}$ (i.e., $\widetilde{\mathbf{a}}_s$, $s = 1, \ldots, S$),
the joint transmit beamforming and ideal phase-shift optimization problem for the $s$-th BS can be expressed as
\begin{subequations}
\label{eq:PM problem for each BS}
\begin{align}\label{eq:PM problem for each BS a}
   \min\limits_{{\mathbf{W}}_s, {\bm{\phi}}_s}~~&
   \sum_{k_s\in\mathcal{K}_s}\left\|{\mathbf{w}}_{s,k_s}\right\|^2\\
   \label{eq:PM problem for each BS b}
   \textrm{s.t.}~~& \gamma_{s,k_s} \geq \Gamma_{s,k_s}, \;\;\forall k_s, \\
   \label{eq:PM problem for each BS c}
   &{\bm{\theta}}_s = \exp\left(\jmath\bm{\phi}_s\odot \widetilde{\mathbf{a}}_s\right), \\ \label{eq:PM problem for each BS d}
   &\phi_{s,m} \in (0, 2\pi], \;\;\forall m.
\end{align}
\end{subequations}
Before solving this problem, we present a sufficient condition for the feasibility of problem (\ref{eq:PM problem for each BS}).
If the equivalent channels $\mathbf{G}_s^H\mathbf{H}_{\text{r},s} + \mathbf{H}_{\text{d},s}, \forall s$ are full rank, where
$\mathbf{H}_{\mathrm{d},s} \triangleq
\left[\mathbf{h}_{\mathrm{d}, s, 1}, \cdots, \mathbf{h}_{\mathrm{d}, s, K_s}\right] \in \mathbb{C}^{N_\text{t} \times K_s}$
and $\mathbf{H}_{\mathrm{r},s} \triangleq
\left[\mathbf{h}_{\mathrm{r}, s, 1}, \cdots, \mathbf{h}_{\mathrm{r}, s, K_s}\right] \in \mathbb{C}^{M \times K_s}$,
then problem (\ref{eq:PM problem for each BS}) is feasible for any finite QoS requirement $\Gamma_{k_s}$.
The specific proof is shown in \textit{Proposition 1} in \cite{Two-stage}. Note that this full rank assumption has been widely adopted in the literature \cite{Hongyu}, \cite{IRS_discrete}, \cite{IRS_UCsi}. In fact, the channel coefficient matrix always satisfies full rank, when it follows some continuous independently identical distributions (i.i.d.), e.g. Rayleigh or Ricing. Therefore, in this paper, we also assume that the equivalent channels are full rank and consequently problem (\ref{eq:PM problem for each BS}) is feasible.
Moreover, some special cases that any equivalent channel is low rank, the problem can also be handled by finding a suitable initial point or the method in \cite{feasibility}.

Notice that problem (\ref{eq:PM problem for each BS}) is a non-convex problem due to the coupled variables in constraint (\ref{eq:PM problem for each BS b}) and the non-convex reflection constraint (\ref{eq:PM problem for each BS c}).
Therefore, we utilize the BCD method to iteratively solve for $\mathbf{W}_s$ and $\bm{\phi}_s$, which is described in the rest of this subsection.

\newcounter{TempEqCnt}
\setcounter{TempEqCnt}{\value{equation}}
\setcounter{equation}{11}
\begin{figure*}[t]
\begin{subequations}
\label{eq:Ds bs}
\begin{align}
\mathbf{D}_s &\triangleq \sum_{k_s \in \mathcal{K}_s}\Big(\widehat{\mathbf{d}}_{s,k_s,k_s} \widehat{\mathbf{d}}_{s,k_s,k_s}^H-\Gamma_{s,k_s,k_s}\sum_{j \in \mathcal{K}_s \atop j \neq k_s}\widehat{\mathbf{d}}_{s,k_s,j} \widehat{\mathbf{d}}_{s,k_s,j}^H\Big),\\
\mathbf{b}_s &\triangleq \sum_{k_s \in \mathcal{K}_s}\Big[\big(b_{s,k_s,k_s}^*+\mathbf{1}^H\overline{\mathbf{d}}_{s,k_s,k_s}^*\big)\mathbf{d}_{s,k_s,k_s} - \Gamma_{s,k_s}\sum_{j \in \mathcal{K}_s \atop j \neq k_s} \big(b_{s,k_s,j}^*+\mathbf{1}^H\overline{\mathbf{d}}_{s,k_s,j}^*\big)\mathbf{d}_{s,k_s,j}\Big].
\end{align}\end{subequations}
\rule[-0pt]{18.5 cm}{0.05em}\vspace{-0.2 cm}
\end{figure*}
\setcounter{equation}{\value{TempEqCnt}}

\textit{1) Update $\mathbf{W}_s$}:
Given the service selection matrix $\mathbf{A}$ and the ideal phase-shift vector  $\bm{\phi}_s$, the practical IRS reflection vector $\bm{\theta}_s$ is fixed.
Thus, the combined effective channel from the $s$-th BS to the $k_s$-th user is determined as $\mathbf{h}_{s,k_s}^H \triangleq {\mathbf{h}^{H}_{\text{r},s,k_s}} {\bm{\Theta}}_{s}  \mathbf{G}_{s} + \mathbf{h}_{\text{d},s,k_s}^H$.
Then, the sub-problem for optimizing $\mathbf{W}_s$ is formulated as
\begin{subequations}
    \label{eq:PM problem for Ws}
   \begin{align}
    \min\limits_{{\mathbf{W}}_s}  ~~&
    \sum_{k_s\in\mathcal{K}_s}||{\mathbf{w}}_{s,k_s}||^2\\
        \textrm{s.t.}~~& \frac{|\mathbf{h}_{s,k_s}^H{\mathbf{w}}_{s,k_s}|^2}{\sum^{j\in\mathcal{K}_s}_{j \neq k_s}|\mathbf{h}_{s,k_s}^H{\mathbf{w}}_{s,j}|^2+\sigma^2} \geq \Gamma_{s,k_s}, \;\;\forall k_s,
    \end{align}
\end{subequations}
which is a standard SOCP problem and can be easily solved using the popular convex optimization toolbox such as CVX \cite{CVX}.

\textit{2) Update $\bm{\phi}_s$}:
After obtaining the transmit beamforming $\mathbf{W}_s$, the objective of the power minimization problem (\ref{eq:PM problem for each BS}) is determined, which makes the IRS phase-shift design as a feasibility-check problem with lots of possible solutions, which may not guarantee the converge of the iteration algorithm. Therefore, another proper objective function with respect to $\bm{\phi}_s$ is required to accelerate the convergence.
A widely used method is forcing the QoS constraint (\ref{eq:PM problem for each BS b}) to be more stricter to provide additional DoFs for the power minimization in the next iteration.

For brevity, we first define
\setcounter{equation}{12}
\begin{subequations}
\begin{align}
\mathbf{d}_{s,k_s,j} &\triangleq \text{diag}(\mathbf{h}^{H}_{\text{r},s,k_s}) \mathbf{G}_{s}  {\mathbf{w}}_{s,j},\\
b_{s,k_s,j}&\triangleq \mathbf{h}^{H}_{\text{d},s,k_s}{\mathbf{w}}_{s,j},
\end{align}
\end{subequations}
and reformulate the QoS constraint (\ref{eq:PM problem for each BS b}) as
\begin{equation} \label{eq:QoS constraint}
\begin{aligned}
\big|\bm{\theta}_s^T&\mathbf{d}_{s,k_s,k_s}+b_{s,k_s,k_s}\big|^2\\
&-\Gamma_{s,k_s}\big(\sum_{j \in \mathcal{K}_s \atop j \neq k_s}
\big|\bm{\theta}_s^T\mathbf{d}_{s,k_s,j}+b_{s,k_s,j}\big|^2+\sigma^2\big)
\geq 0.
\end{aligned}
\end{equation}
Then, the optimization problem for $\bm{\phi}_s$ is formulated to maximize the sum of the left-hand side of (\ref{eq:QoS constraint}) as
\begin{subequations}
\label{eq:PM problem theta max sum SINR}
\begin{align}\label{eq:PM problem theta max sum SINR a}
   \max\limits_{{\bm{\phi}}_s}&
   \sum_{k_s\in\mathcal{K}_s}\Big( \big|\bm{\theta}_s^T\mathbf{d}_{s,k_s,k_s}+b_{s,k_s,k_s}\big|^2\nonumber\\
   &~~~~~~~~~~~~
   -\Gamma_{s,k_s}\hspace{-0.1 cm}\sum_{j \in \mathcal{K}_s \atop j \neq k_s}\big|\bm{\theta}_s^T\mathbf{d}_{s,k_s,j}+b_{s,k_s,j}\big|^2\Big) \\
   \label{eq:PM problem theta max sum SINR b}
   ~~\text{s.t.}&~~{\bm{\theta}}_s = \exp\left(\jmath\bm{\phi}_s\odot \widetilde{\mathbf{a}}_s\right), \\
   \label{eq:PM problem theta max sum SINR c}
   &~~\phi_{s,m} \in (0, 2\pi], \;\;\forall m.
\end{align}
\end{subequations}
Different from the IRS reflection designs with the ideal reflection model, we notice that the practical reflection coefficient $\theta_{s,m}$ is determined by both the ideal phase-shift $\phi_{s,m}$ and the service selection indicator $a_{s,m}$ in constraint (\ref{eq:PM problem theta max sum SINR b}).
Particularly, when $a_{s,m} = 0$, $\theta_{s,m}$ is fixed as 1 regardless of the value $\phi_{s,m}$, i.e., optimizing $\phi_{s,m}$ cannot influence the objective value (\ref{eq:PM problem theta max sum SINR a}).
Therefore, we only need to optimize the phase-shift vector of the selected reflecting elements.
Following this analysis, we divide the set of reflecting elements $\mathcal{M}$ into the set of selected reflecting elements $\mathcal{I}_s \triangleq \{m|a_{s,m} = 1\}$ and the set of others $\mathcal{E}_s \triangleq \{m|a_{s,m} = 0\}$, and then correspondingly partition the phase-shift vector $\bm{\phi}_s$ into $\widehat{\bm{\phi}}_s$ which contains $\phi_{s,m}, \forall m\in\mathcal{I}_s$ and $\overline{\bm{\phi}}_s$ which consists of $\phi_{s,m}, \forall m\in\mathcal{E}_s$, the practical reflection $\bm{\theta}_s$ into $\widehat{\bm{\theta}}_s$ and $\overline{\bm{\theta}}_s$, and $\mathbf{d}_{s,k_s,j}$ into $\widehat{\mathbf{d}}_{s,k_s,j}$ and $\overline{\mathbf{d}}_{s,k_s,j}$.
It is obvious that $\widehat{\bm{\phi}}_s$ is the required optimization variable, $\overline{\bm{\theta}}_s = \mathbf{1}$ and constraint (\ref{eq:PM problem theta max sum SINR b}) is converted to $\widehat{\bm{\theta}}_s = \exp(\jmath\widehat{\bm{\phi}}_s)$.
Thus, $\widehat{\bm{\phi}}_s$ can be obtained by taking the angle of $\widehat{\bm{\theta}}_s$, and problem (\ref{eq:PM problem theta max sum SINR}) with respect to $\widehat{\bm{\theta}}_s$ can be reformulated as
\begin{subequations}\label{eq:PM problem theta}
\begin{align}\label{eq:PM problem theta a}
\min\limits_{\widehat{\bm{\theta}}_s}~~ &-\widehat{\bm{\theta}}_s^T\mathbf{D}_s\widehat{\bm{\theta}}^*_s - \widehat{\bm{\theta}}_s^T\mathbf{b}_s - \mathbf{b}_s^H\widehat{\bm{\theta}}_s \\
\label{eq:PM problem theta b}
\textrm{s.t.}~~&|\widehat{\theta}_{s,m}| = 1,\;\;m = 1, \ldots, \left|\mathcal{I}_s\right|,
\end{align}
\end{subequations}
where $\mathbf{D}_s$ and $\mathbf{b}_s$ are defined in (\ref{eq:Ds bs}) as shown at the top of this page.

It can be observed that the main difficulty to tackle problem (\ref{eq:PM problem theta}) is the non-convex unit modulus constraint (\ref{eq:PM problem theta b}).
In the literature of IRS optimization, non-convex relaxation, e.g., SDP/SDR, MM, and alternating minimization methods are usually exploited to handle constraint (\ref{eq:PM problem theta b}) on the Euclidean space.
However, a fast convergence and satisfactory performance cannot be guaranteed since either the objective function or the constraint is relaxed using these methods.
Therefore, we propose to solve problem (\ref{eq:PM problem theta}) on the Riemannian space.
Specifically, the unit modulus constraint (\ref{eq:PM problem theta b}) forms a smooth Riemannian manifold, on which problem (\ref{eq:PM problem theta}) becomes an unconstrained problem.
Since the Riemannian manifold resembles a Euclidean space at each point, lots of classic algorithms developed on the Euclidean space, e.g., steepest descent, conjugate gradient, quasi-Newton methods, have found their counterparts on the Riemannian manifold \cite{IRS_Liuxing2}, \cite{Liuxing}.
In addition, since the objective function (\ref{eq:PM problem theta a}) is a typical quadratic function with respect to $\widehat{\bm{\theta}}_s$,  the required first-order and second-order derivatives in the above methods can be easily calculated as $-2\mathbf{D}_s\widehat{\bm{\theta}}^*_s - 2\mathbf{b}_s^H$ and $-2\mathbf{D}_s$, respectively.
Therefore, problem (\ref{eq:PM problem theta}) can be readily solved using the advanced Riemannian manifold based algorithms.
Due to space limitations, the details are omitted in this paper.

After obtaining the solution $\widehat{\bm{\theta}}_s^\star$ of problem (\ref{eq:PM problem theta}), the ideal phase-shift $\phi_{s,m}^\star, \forall m$, is updated by
\begin{equation}\label{eq:construct phi}
\phi_{s,m}^\star = \left\{
             \begin{array}{lr}
             \angle\widehat{\theta}_{s,m}^\star, &a_{s,m} = 1, \\
             \phi_{s,m}, &a_{s,m} = 0.
             \end{array}
\right.
\end{equation}
\vspace{-0.9 cm}

\subsection{Service Selection Design}
With obtained transmit beamforming $\mathbf{W}_s$ and ideal phase-shift vector $\bm{\phi}_s$ for each BS, the original problem (\ref{eq:PM problem}) is transformed into a service selection design problem.
Since the service selection matrix $\mathbf{A}$ cannot directly affect (\ref{eq:PM problem a}), we should formulate another optimization problem to achieve better QoS for all users excepting to provide a larger feasible area, which can assist the joint transmit beamforming and practical phase-shift design in the next stage.
Thus, similar to problem (\ref{eq:PM problem theta max sum SINR}), the optimization problem for updating $\mathbf{A}$ is formulated as
\begin{subequations}
\label{eq:PM problem A}
\begin{align}\label{eq:PM problem A a}
   \max\limits_{\mathbf{A}}&~
   \sum_{s\in\mathcal{S}}
   \sum_{k_s\in\mathcal{K}_s}\Big( \big|\bm{\theta}_s^T\mathbf{d}_{s,k_s,k_s}+b_{s,k_s,k_s}\big|^2 \nonumber\\
   &\hspace{1.8 cm} -\Gamma_{s,k_s}\sum_{j \in \mathcal{K}_s \atop j \neq k_s}\big|\bm{\theta}_s^T\mathbf{d}_{s,k_s,j}+b_{s,k_s,j}\big|^2\Big) \\
   \label{eq:PM problem A b}
   ~\text{s.t.}&~~{\bm{\theta}}_s = \exp\left(\jmath\bm{\phi}_s\odot \widetilde{\mathbf{a}}_s\right),\;\;\forall s,\\
   \label{eq:PM problem A c}
   &~\left\| \mathbf{a}_m \right\|_0 \leq 1, \;\;\forall m,\\
   \label{eq:PM problem A d}
   &~~a_{s,m} \in \{0,1\}, \;\;\forall s,\; m.
    \end{align}
\end{subequations}
Since the objective function is implicit with the variable $\mathbf{A}$, we first re-write it by substituting the equality constraint (\ref{eq:PM problem A b}) into (\ref{eq:PM problem A a}).
Thus, the term $\bm{\theta}_s^T\mathbf{d}_{s,k_s,j}+b_{s,k_s,j}$ in (\ref{eq:PM problem A}a) can be equivalently re-written as
\begin{subequations}\label{eq:the term in obj}
\begin{align}
&\bm{\theta}_s^T\mathbf{d}_{s,k_s,j}+b_{s,k_s,j} \non \\
&=\big(\widetilde{\mathbf{a}}_s^T \odot e^{\jmath\bm{\phi}_s^T}\big)
        \mathbf{d}_{s,k_s,j} + (\bm{1} -{\widetilde{\mathbf{a}}_s})^T\mathbf{d}_{s,k_s,j}+b_{s,k_s,j}\\
&=\widetilde{\mathbf{a}}_s^T\text{diag}\{e^{\jmath\bm{\phi}_s} -\bm{1}\}\mathbf{d}_{s,k_s,j}+\bm{1}^T \mathbf{d}_{s,k_s,j} +b_{s,k_s,j} \\
&=\widetilde{\mathbf{a}}_s^T\widetilde{\mathbf{d}}_{s,k_s,j}+\widetilde{b}_{s,k_s,j},
\end{align}
\end{subequations}
where for simplicity we define
\begin{subequations}
\begin{align}
\widetilde{\mathbf{d}}_{s,k_s,j} &\triangleq \mathrm{diag}
(e^{\jmath\bm{\phi}_s} -\bm{1})\mathbf{d}_{s,k_s,j},\;\; \forall s,\;k_s,\;j,\\
\widetilde{b}_{s,k_s,j} &\triangleq \bm{1}^T\mathbf{d}_{s,k_s,j} + {b}_{s,k_s,j}, \;\;\forall s,\;k_s,\;j.
\end{align}
\end{subequations}
Then, by plugging (\ref{eq:the term in obj}c) into the objective (\ref{eq:PM problem A a}), expanding the quadratic terms, and ignoring the constant terms, the original problem (\ref{eq:PM problem A}) can be equivalently re-formulated as
\begin{subequations}
\label{eq:PM problem A2}
\begin{align}\label{eq:PM problem A2 a}
   \max\limits_{\mathbf{A}}&~
   \sum_{s\in\mathcal{S}}
   \widetilde{\mathbf{a}}_s^T\widetilde{\mathbf{E}}_{s}\widetilde{\mathbf{a}}_s
   -\sum_{s\in\mathcal{S}}
   \widetilde{\mathbf{a}}_s^T\widetilde{\mathbf{D}}_{s}\widetilde{\mathbf{a}}_s
   +\sum_{s\in\mathcal{S}}
   \mathfrak{R}\{\widetilde{\mathbf{a}}_s^T\widetilde{\bm{\beta}}_s\}\\
   \text{s.t.}&~
   \label{eq:PM problem A2 b}
   \left\| \mathbf{a}_m \right\|_0 \leq 1, \;\;\forall m,\\
   \label{eq:PM problem A2 c}
   &~~a_{s,m} \in \{0,1\}, \;\;\forall s,\; m,
\end{align}
\end{subequations}
where we define
\begin{subequations}
\begin{align}
    \widetilde{\mathbf{E}}_{s} \triangleq&~
    \sum_{k_s\in\mathcal{K}_s}(1+\Gamma_{s,k_s})
    \mathbf{d}_{s,k_s,j = k_s}\mathbf{d}^H_{s,k_s,j = k_s},
    \;\;\forall s,\\
    \widetilde{\mathbf{D}}_{s} \triangleq&~
    \sum_{k_s\in\mathcal{K}_s}\Gamma_{s,k_s}\sum_{j\in\mathcal{K}_s}
    \mathbf{d}_{s,k_s,j}\mathbf{d}^H_{s,k_s,j},
    \;\;\forall s,\\
    \widetilde{\bm{\beta}}_{s} \triangleq&~
    2\sum_{k_s\in\mathcal{K}_s}\big(
    \mathbf{d}_{s,k_s,j = k_s}b^*_{s,k_s,j = k_s}\nonumber\\
    &\hspace{1.8cm}
    -\Gamma_{s,k_s}\sum_{j\neq k_s}\mathbf{d}_{s,k_s,j}b^*_{s,k_s,j}
    \big), \;\;\forall s.
\end{align}
\end{subequations}
It can be observed that problem (\ref{eq:PM problem A2}) cannot be directly solved due to the non-convex objective function (\ref{eq:PM problem A2}a) and the non-smooth constraints (\ref{eq:PM problem A2}b) and (\ref{eq:PM problem A2}c).
Therefore, we propose to transform both the constraints and the objective function into more tractable forms in the followings.

\begin{algorithm}[t]\small
\caption{Joint Transmit Beamforming and Practical IRS Reflection Design for Power Minimization Problem}
\label{alg:SH}
    \begin{algorithmic}[1]
    \REQUIRE $\mathbf{h}_{\text{r},s,k_s}^H$, $\mathbf{G}_s$, $\mathbf{h}_{\text{d},s,k_s}^H$, $\Gamma_{s,k_s}$, $\forall s \in \mathcal{S}$, $\forall k_s \in \mathcal{K}_s$, $\sigma^2$.
    \ENSURE $\mathbf{W}_s^\star$, $\bm{\phi}_s^\star$, $\forall s\in\mathcal{S}$,\; $\mathbf{A}^\star$.
    \STATE {Initialize $\mathbf{A} = \mathbf{1}$, $\bm{\Phi}$.}
        \FOR {$s = 1$ to $S$}
            \WHILE{no convergence for (\ref{eq:PM problem for each BS a})}
                \STATE{Calculate $\mathbf{W}_{s}^{\star}$ by solving (\ref{eq:PM problem for Ws}) with CVX.}
                \STATE {Obtain $\widehat{\bm{\theta}}_s^\star$ by solving  (\ref{eq:PM problem theta}) with Riemannian manifold optimizations.}
                \STATE{Construct ideal phase-shift $\bm{\phi}_s^\star$ by (\ref{eq:construct phi}).}
            \ENDWHILE
        \ENDFOR
        \WHILE{no convergence for $\mathbf{A}$}
            \STATE{Obtain $\mathbf{A}$ by solving (\ref{eq:problem AA}).}
        \ENDWHILE
    \STATE {Repeat steps 2-8.}
    \STATE {Return $\mathbf{W}_s^\star$,\;$\bm{\phi}_s^\star$, $\forall s\in\mathcal{S}$,\;$\mathbf{A}^\star = [\mathbf{a}_1^\star,\ldots,\mathbf{a}_M^\star]$.}
    \end{algorithmic}
\end{algorithm}

Considering that each element of $\mathbf{A}$ is binary as shown in constraint (\ref{eq:PM problem A2}c), the 0-norm constraint (\ref{eq:PM problem A2}b) can be equivalently transformed into a 1-norm constraint and re-formulated as:
\begin{equation}
||\mathbf{a}_m||_0 = ||\mathbf{a}_m||_1 = \mathbf{1}^T\mathbf{a}_m \leq 1, \;\;\forall m.
\end{equation}
Then, by utilizing the difference-of-convex function method in \cite{DC}, the binary constraint (\ref{eq:PM problem A2}c) is be transformed into:
\begin{subequations}\label{eq:transform binary constraint}
\begin{align}
& 0 \leq a_{s,m} \leq 1, \;\;\forall s,\; m,\\
\label{eq:DC non-convex}
& \sum_{s \in \mathcal{S}}\big(
\mathbf{1}^T\widetilde{\mathbf{a}}_{s}-\widetilde{\mathbf{a}}^T_{s}\widetilde{\mathbf{a}}_{s}\big) \leq 0.
\end{align}
\end{subequations}
Since the constraint (\ref{eq:transform binary constraint}b) is non-convex, we further apply the penalty method to move it to the objective and re-formulate the optimization problem as
\begin{subequations}
\label{eq:PM problem A3}
\begin{align}
\label{eq:PM problem A3 a}
\max\limits_{\mathbf{A}}&~
   \sum_{s\in\mathcal{S}}
   \widetilde{\mathbf{a}}_s^T\widetilde{\mathbf{E}}_{s}\widetilde{\mathbf{a}}_s
   -\sum_{s\in\mathcal{S}}
   \widetilde{\mathbf{a}}_s^T\widetilde{\mathbf{D}}_{s}\widetilde{\mathbf{a}}_s
   +\sum_{s\in\mathcal{S}}   \mathfrak{R}\{\widetilde{\mathbf{a}}_s^T\widetilde{\bm{\beta}}_s\}\nonumber\\
&~~ -\tau\sum_{s \in \mathcal{S}}\big(
\mathbf{1}^T\widetilde{\mathbf{a}}_{s}-\widetilde{\mathbf{a}}^T_{s}\widetilde{\mathbf{a}}_{s}\big)\\  ~\text{s.t.}
&~~ \mathbf{1}^T\mathbf{a}_m \leq 1, \;\;\forall m,\\
&~~  0 \leq a_{s,m} \leq 1, \;\;\forall s,\; m,
\end{align}
\end{subequations}
where $\tau > 0$ is a penalty factor.
It is proved in \cite{plenty} that the transform is equivalent when $\tau$ has a moderately high value.
After obtaining the preferable linear constraints (\ref{eq:PM problem A3}b) and (\ref{eq:PM problem A3}c), the non-convex objective function (\ref{eq:PM problem A3}a) is the major obstacle.
In order to efficiently solve this problem, we employ the MM method and seek for a linear surrogate function of (\ref{eq:PM problem A3}a) by utilizing the first-order Taylor expansion.
Specifically, a lower-bounded surrogate function of the first term and last term in (\ref{eq:PM problem A3}a) in the $t$-th iteration is derived as
\begin{subequations}\label{eq:MM derivation}
\begin{align}
&\sum_{s\in\mathcal{S}}
   \widetilde{\mathbf{a}}_s^T\widetilde{\mathbf{E}}_{s}\widetilde{\mathbf{a}}_s
    -\tau\sum_{s \in \mathcal{S}}\big(
\mathbf{1}^T\widetilde{\mathbf{a}}_{s}-\widetilde{\mathbf{a}}^T_{s}\widetilde{\mathbf{a}}_{s}\big)\\
\label{eq:MM derivation b}
&\geq\sum_{s\in\mathcal{S}}\big[
   ({\widetilde{\mathbf{a}}_s^t)}^T\widetilde{\mathbf{E}}_{s}\widetilde{\mathbf{a}}_s^t
   + 2\mathfrak{R}\big\{(\widetilde{\mathbf{a}}_{s} - \widetilde{\mathbf{a}}_s^t)^T
   \widetilde{\mathbf{E}}_{s}\widetilde{\mathbf{a}}_s^t\big\}\big] \nonumber\\
&~~~~  - \tau\sum_{s\in\mathcal{S}}\big[\mathbf{1}^T\widetilde{\mathbf{a}}_{s}
       -{(\widetilde{\mathbf{a}}^t_{s})}^T\widetilde{\mathbf{a}}^t_{s}
       -2{(\widetilde{\mathbf{a}}^t_{s})}^T(\widetilde{\mathbf{a}}_{s}-\widetilde{\mathbf{a}}^t_{s})
       \big],
\end{align}
\end{subequations}
where the vector $\widetilde{\mathbf{a}}^t_{s}$ denotes the solution of the last iteration.

Based on the above derivations, by substituting the surrogate function (\ref{eq:MM derivation b}) into (\ref{eq:PM problem A3}a), the optimization problem in the $t$-th iteration can be formulated as
\begin{subequations}
\label{eq:problem AA}
\begin{align}
\max\limits_{\mathbf{A}}&~~-\sum_{s\in\mathcal{S}}
   \widetilde{\mathbf{a}}_s^T\widetilde{\mathbf{D}}_{s}\widetilde{\mathbf{a}}_s
    + \sum_{s\in\mathcal{S}}\mathfrak{R}\{(\widetilde{\mathbf{a}}_s^T\bm{\beta}_s\} + c\\
~\text{s.t.}
&~~ \mathbf{1}^T\mathbf{a}_m \leq 1, \;\;\forall m,\\
&~~  0 \leq a_{s,m} \leq 1, \;\;\forall s,\; m.
\end{align}
\end{subequations}
where
\begin{subequations}\label{eq:betac}
\begin{align}
\bm{\beta}_s \triangleq &~
2\widetilde{\mathbf{E}}_{s}\widetilde{\mathbf{a}}^t_{s}+\tau\widetilde{\mathbf{a}}^t_{s} +\widetilde{\bm{\beta}}_{s}, \;\;\forall s,\\
c \triangleq &~
\sum_{s\in\mathcal{S}}\big[
   ({\widetilde{\mathbf{a}}_s^t)}^T\widetilde{\mathbf{E}}_{s}\widetilde{\mathbf{a}}_s^t
   - 2\mathfrak{R}\{(\widetilde{\mathbf{a}}_s^t)^T
   \widetilde{\mathbf{E}}_{s}\widetilde{\mathbf{a}}_s^t\}\nonumber\\
   &~~~~~
   - (\widetilde{\mathbf{a}}_s^t)^T\lambda_s\widetilde{\mathbf{a}}_s^t
 - \tau{(\widetilde{\mathbf{a}}^t_{s})}^T\widetilde{\mathbf{a}}^t_{s}\big].
\end{align}
\end{subequations}
It is obvious that problem (\ref{eq:problem AA}) is a convex problem that can be efficiently solved by various existing methods.

\subsection{Summary and Complexity Analysis}
\textit{Summary}:
Based on the above derivations, the joint transmit beamforming and IRS reflection design for the power minimization problem (\ref{eq:PM problem}) is straightforward and summarized in Algorithm 1.
Given initialization $\mathbf{A} = \mathbf{1}$, the transmit beamforming and ideal phase-shifts for each BS are iteratively updated.
Note that the reformulated objective function (13) is utilized in the iteration, theoretical convergence analysis cannot be easily obtained. But the simulation results shown in Sec. V illustrate that the proposed algorithm will converge fast to a local optimum.
Then, the service selection matrix $\mathbf{A}^\star$ is optimized based on the obtained transmit beamforming and ideal phase-shift results.
Finally, $\mathbf{W}_s^\star$ and $\bm{\phi}_s^\star$ are solved one more time with the obtained service selection matrix $\mathbf{A}^\star$.

\textit{Complexity analysis}:
In each iteration, updating the transmit beamforming $\mathbf{W}_s$ for each BS by solving a SOCP problem has a complexity of $\mathcal{O}(N_\mathrm{t}^{4.5}K_s^{3.5})$; updating the ideal phase-shift vector $\widehat{\bm{\theta}}_s$ using Riemannian manifold optimizations has a complexity of $\mathcal{O}(M^{1.5})$ at most.
The complexity for updating the service selection matrix $\mathbf{A}$ is
$\mathcal{O}((SM)^{3})$.
Therefore, the total complexity of the proposed algorithm is of order
$\mathcal{O}\{\sum_{s=1}^S(2N_\mathrm{t}^{4.5}K_s^{3.5}+2M^{1.5})+ (SM)^{3}\}$.

\section{Algorithms for Sum-rate Maximization Problem}
In this section, we investigate the sum-rate maximization problem in the considered multi-cell multi-band systems.
It is worth noting that the power minimization problem and the sum-rate maximization problem cannot be directly converted from one to another.
In order to effectively handle this non-convex NP-hard optimization problem, the WMMSE approach and BCD method are employed to iteratively solve for each variable with a closed-form solution.

Specifically, our goal is to jointly optimize the transmit beamformers $\mathbf{W}$ for the users of all BSs, the ideal IRS phase-shifts $\bm{\Phi}$, and the service selection matrix $\mathbf{A}$ to maximize the sum-rate, subject to the transmit power budget of each BS and the practical IRS reflection model.
Therefore, the optimization problem is formulated as
\begin{subequations}
 \label{eq:sum rate problem}
   \begin{align} \label{eq:sum rate problem a}
   \max\limits_{\mathbf{W},\bm{\Phi},\mathbf{A}}~~&
   \sum_{s\in\mathcal{S}} \sum_{k_s\in\mathcal{K}_s}\log_2(1+\gamma_{s,k_s})\\
   \label{eq:sum rate problem b}
   \textrm{s.t.}~~&\sum_{k_s\in\mathcal{K}_s}\left\|\mathbf{w}_{s,k_s}\right\|^2 \leq P_s,\;\;\forall s,\\
    \label{eq:sum rate problem c}
    &\bm{\theta}_s = \exp\left(\jmath\bm{\phi}_s \odot \widetilde{\mathbf{a}}_s\right),\;\; \forall s, \\
   &\phi_{s,m} \in (0, 2\pi], \;\;\forall s,\; m, \\
    \label{eq:sum rate problem e}
   &\left\| \mathbf{a}_m \right\|_0 \leq 1, \;\;a_{s,m} \in \{0,1\}, \;\;\forall s,\; m,
\end{align}
\end{subequations}
where $P_s > 0$ denotes the transmit power budget of the $s$-th BS.
Seeking for the solution to this non-convex NP-hard problem is very difficult, not only due to the complicated objective function (\ref{eq:sum rate problem a}) that contains the fractional terms in $\log({\cdot})$, but also because of the coupled variables in the objective function (\ref{eq:sum rate problem a}) and constraints (\ref{eq:sum rate problem c}).
Therefore, in the followings we first employ the WMMSE approach to convert the original optimization problem (\ref{eq:sum rate problem}) into a more tractable multi-variate problem and then use the BCD method to iteratively solve for each variable.

Although the WMMSE approach has been widely used for solving the sum-rate maximization problem, we would like to emphasize that the IRS beamforming design with the practical reflection model will be quite different and difficult due to the complicated constraint of each phase-shift element. Therefore, we need to develop an efficient algorithm to solve this sum-rate maximization problem (\ref{eq:sum rate problem}), which for the first time considers the frequency-selective characteristic of the IRS in multi-cell networks.

\subsection{Problem Reformulation by WMMSE}
Following the derivations in \cite{WMMSE}, a scalar $\nu_{s,k_s} \in \mathbb{C}$ is applied for the $k_s$-th user to estimate the transmitted signal $z_{s,k_s}$.
Then, the MSE of the $k_s$-th user can be calculated as
\begin{equation}\label{eq:MSE}
\begin{aligned}
   \text{MSE}_{s,k_s} \hspace{-0.1cm} = & \mathbb{E}\big\{(\nu_{s,k_s}^* y_{s,k_s}-z_{s,k_s} )(\nu_{s,k_s}^* y_{s,k_s}-z_{s,k_s} )^* \big\}  \\
   =& \sum_{j\in\mathcal{K}_s}\big|\nu_{s,k_s}^*
   ({\mathbf{h}^{H}_{\text{r},s,k_s}}
    \bm{\Theta}_s
    \mathbf{G}_{s} + \mathbf{h}_{\text{d},s,k_s}^H)
    \mathbf{w}_{s,j}\big|^2  \\
    & -2\mathfrak{R}\big\{\nu_{s,k_s}^*
   ({\mathbf{h}^{H}_{\text{r},s,k_s}}
    \bm{\Theta}_s
    \mathbf{G}_{s} + \mathbf{h}_{\text{d},s,k_s}^H)
    \mathbf{w}_{s,k_s}\big\}\\
    & +|\nu_{s,k_s}|^2\sigma^2 + 1,\;\;\forall s,\;k_s.
\end{aligned}
\end{equation}
Introducing the MSE weights $\mu_{s,k_s} \in \mathbb{R}^+, \forall s, k_s$, the sum-rate maximization problem (\ref{eq:sum rate problem}) is equivalently reformulated as:
\begin{subequations}
\label{eq:sum rate problem reformualte}
\begin{align}\label{eq:sum rate problem reformualte a}
\min \limits_{\mathbf{W},\bm{\Phi},\mathbf{A},\bm{\nu},\bm{\mu}}~~ &\sum_{s\in\mathcal{S}}\sum_{k_s\in\mathcal{K}_s}
\left(\mu_{s,k_s}\text{MSE}_{s,k_s}-\log_2\mu_{s,k_s}\right) \\
\textrm{s.t.}~~& (\text{\ref{eq:sum rate problem b}})-(\text{\ref{eq:sum rate problem e}}),
\end{align}
\end{subequations}
where $\bm{\mu}$ and $\bm{\nu}$ denote the vectors that contain the auxiliary variables $\mu_{s,k_s}$ and $\nu_{s,k_s},\forall s,k_s$, respectively.
Since the complicated $\log_2(1+\gamma_{s,k_s})$ term in the objective function is transformed into a polynomial term and a very simple $\log_2(\cdot)$ term, the reformulated optimization problem (\ref{eq:sum rate problem reformualte}) is much more tractable.
Specifically, the obtained multi-variate optimization problem (\ref{eq:sum rate problem reformualte}) can be solved using the typical BCD method.
The details for updating each variable are presented in the next subsection.

\setcounter{TempEqCnt}{\value{equation}}
\setcounter{equation}{33}
\begin{figure*}[t]
\begin{subequations}\label{eq:Bs cs}
\begin{align}
\mathbf{B}_s &\triangleq \sum_{k_s\in\mathcal{K}_s} \mu_{s,k_s}|\nu_{s,k_s}|^2
\text{diag}\{\mathbf{h}_{\text{r},s,k_s}^H\}\mathbf{G}_s\sum_{j\in\mathcal{K}_s}
\mathbf{w}_{s,j}\mathbf{w}_{s,j}^H\mathbf{G}_s^H\text{diag}\{\mathbf{h}_{\text{r},s,k_s}\}, \forall s, \\
    \mathbf{c}_s &\triangleq \sum_{k_s\in\mathcal{K}_s}\mu_{s,k_s}\nu_{s,k_s}^*\text{diag}\{\mathbf{h}_{\text{r},s,k_s}^H\}
    \mathbf{G}_s\mathbf{w}_{s,k_s}\Big(1-\nu_{s,k_s}\sum_{j\in\mathcal{K}_s}
    \mathbf{w}_{s,j}\mathbf{w}_{s,j}^H\mathbf{h}_{\text{d},s,k_s}\Big), \forall s.
\end{align}
\end{subequations}
\hrulefill
\end{figure*}
\setcounter{equation}{\value{TempEqCnt}}

\subsection{Block Update by BCD}
\subsubsection{Update $\bm{\nu}$}
With given the transmit beamformer $\mathbf{W}$, the ideal IRS phase-shift $\bm{\Phi}$, the service selection matrix $\mathbf{A}$, and the MSE weight vector $\bm{\mu}$, the design of $\bm{\nu}$ for each user is independent.
The optimization problem of solving for $\nu_{s,k_s}$ can be expressed as
\begin{equation}
\begin{aligned}
    \min_{\nu_{s,k_s}}~&|\nu_{s,k_s}|^2 \Big(\hspace{-0.1cm}\sum_{j\in\mathcal{K}_s}\hspace{-0.1cm}\big|
   ({\mathbf{h}^{H}_{\text{r},s,k_s}}
    \bm{\Theta}_s
    \mathbf{G}_{s} \hspace{-0.1cm}+\hspace{-0.1cm} \mathbf{h}_{\text{d},s,k_s}^H)
    \mathbf{w}_{s,j}\big|^2 \hspace{-0.1cm}+\hspace{-0.1cm}\sigma^2\Big)
    \\ ~~&-2\mathfrak{R}\big\{\nu_{s,k_s}^*
   ({\mathbf{h}^{H}_{\text{r},s,k_s}}
    \bm{\Theta}_s
    \mathbf{G}_{s} + \mathbf{h}_{\text{d},s,k_s}^H)
    \mathbf{w}_{s,k_s}\big\},
\end{aligned}
\end{equation}
which is an unconstrained quadratic convex problem.
Thus, setting the first-order derivative with respect to $\nu_{s,k_s}$ to zero, the optimal solution $\nu_{s,k_s}^\star$ can be easily calculated as
\begin{equation}\label{eq:optimal nu}
    \nu_{s,k_s}^\star = \frac{({\mathbf{h}^{H}_{\text{r},s,k_s}} \bm{\Theta}_s\mathbf{G}_{s} + \mathbf{h}_{\text{d},s,k_s}^H) \mathbf{w}_{s,k_s}}
    {\sum_{j\in\mathcal{K}_s}\big|({\mathbf{h}^{H}_{\text{r},s,k_s}}\bm{\Theta}_s\mathbf{G}_{s} + \mathbf{h}_{\text{d},s,k_s}^H) \mathbf{w}_{s,j}\big|^2 + \sigma^2}.
\end{equation}

\subsubsection{Update $\bm{\mu}$}

Fixing the transmit beamformer $\mathbf{W}$, the ideal IRS phase-shift $\bm{\Phi}$, the service selection matrix $\mathbf{A}$, and the variable $\bm{\nu}$, the optimization problem with respect to each independent MSE weight $\mu_{s,k_s}$ can be formulated as
\setcounter{equation}{34}
\begin{equation}
    \min_{\mu_{s,k_s}}~~\mu_{s,k_s}\text{MSE}_{s,k_s}-\log_2 \mu_{s,k_s}. \label{eq:S_rho}
\end{equation}
Similarly, the optimal solution $\mu_{s,k_s}^\star$ can be obtained by applying the typical first-order optimality condition as
\begin{equation}\label{eq:optimal mu}
    \mu_{s,k_s}^\star = \text{MSE}_{s,k_s}^{-1} \overset{(a)}{=} 1+\gamma_{s,k_s},
\end{equation}
where $(a)$ is derived by substituting the optimal $\nu_{s,k_s}^\star$ in (\ref{eq:optimal nu}) into $\text{MSE}_{s,k_s}$ in (\ref{eq:MSE}).

\subsubsection{Update $\mathbf{W}$}
After obtaining the variable $\bm{\nu}$, the MSE weight $\bm{\mu}$, the ideal IRS phase-shift $\bm{\Phi}$, and the service selection matrix $\mathbf{A}$, the optimization problem for designing the transmit beamformer $\mathbf{W}_s$ for the $s$-th BS can be formulated as
\begin{subequations}
\label{eq:sum rate problem Ws}
\begin{align}
    \min \limits_{\mathbf{W}_s}& \sum_{k_s\in\mathcal{K}_s}\hspace{-0.1cm}\mu_{s,k_s}\Big(\hspace{-0.1cm}\sum_{j\in\mathcal{K}_s}
    |\overline{\mathbf{h}}^H_{s,k_s} \mathbf{w}_{s,j}|^2 \hspace{-0.1cm}-\hspace{-0.1cm}2\mathfrak{R}\{\overline{\mathbf{h}}^H_{s,k_s}\mathbf{w}_{s,k_s}\}\Big)\\
    \label{eq:sum rate problem Ws b}
     \textrm{s.t.}~~&\sum_{k_s\in\mathcal{K}_s}||\mathbf{w}_{s,k_s}||^2 \leq P_s,
\end{align}
\end{subequations}
where we define $\overline{\mathbf{h}}_{s,k_s}^H \triangleq\nu_{s,k_s}^* ({\mathbf{h}^{H}_{\text{r},s,k_s}}\bm{\Theta}_s\mathbf{G}_{s} + \mathbf{h}_{\text{d},s,k_s}^H)$ for brevity.
Note that (\ref{eq:sum rate problem Ws}) is a convex optimization problem, which can be readily solved using standard convex optimization algorithms, e.g., the interior-point algorithm.
In order to reduce the execution time, we employ the typical Lagrange multiplier method to obtain a closed-form solution.
Attaching a Lagrange multiplier $\lambda_s \geq 0$ to the power constraint (\ref{eq:sum rate problem Ws b}), the Lagrange function of problem (\ref{eq:sum rate problem Ws}) can be formulated as
\begin{subequations}
\label{eq:sum rate Ws Lagrange function}
\begin{align}
    &\mathcal{L}(\mathbf{W}_s, \lambda_s) \nonumber\\
    &\triangleq\hspace{-0.1cm}\sum_{k_s\in\mathcal{K}_s}\hspace{-0.1cm}\mu_{s,k_s}\Big(\hspace{-0.1cm}\sum_{j\in\mathcal{K}_s}
    \big|\overline{\mathbf{h}}^H_{s,k_s} \mathbf{w}_{s,j}\big|^2 -2\mathfrak{R}\big\{\overline{\mathbf{h}}^H_{s,k_s}\mathbf{w}_{s,k_s}\big\}\Big)\nonumber\\
    &~~~~+\lambda_s\Big(\sum_{k_s\in\mathcal{K}_s}||\mathbf{w}_{s,k_s}||^2- P_s\Big)\\
    &=\hspace{-0.1cm}\sum_{k_s\in\mathcal{K}_s}\hspace{-0.1cm}\Big(\mathbf{w}^H_{s,k_s}
    \hspace{-0.1cm}\sum_{j\in\mathcal{K}_s}\hspace{-0.1cm}\mu_{s,j}\overline{\mathbf{h}}_{s,j}
    \overline{\mathbf{h}}^H_{s,j}\mathbf{w}_{s,k_s}\hspace{-0.1cm}-\hspace{-0.1cm}
    2\mu_{s,k_s}\mathfrak{R}\big\{\overline{\mathbf{h}}_{s,k_s}\mathbf{w}_{s,k_s}\big\}\nonumber\\
      &~~~~+ \lambda_s\mathbf{w}^H_{s,k_s}\mathbf{w}_{s,k_s}\Big) - \lambda_s P_s.
\end{align}
\end{subequations}
Then, setting $\frac{\partial\mathcal{L}(\mathbf{W}_s, \lambda_s)}{\mathbf{w}_{s,k_s}} = \mathbf{0}$, the optimal beamforming vector $\mathbf{w}_{s,k_s}^\star$ is given by
\begin{equation}\label{eq:optimal wsk}
    \mathbf{w}_{s,k_s}^\star =
    \mu_{s,k_s}\Big(\sum_{j\in\mathcal{K}_s}\mu_{s,j}
      \overline{\mathbf{h}}_{s,j}\overline{\mathbf{h}}^H_{s,j} + \lambda_s\mathbf{I}_{N_\text{t}}\Big)^{-1}
\overline{\mathbf{h}}_{s,k_s},
\end{equation}
where the Lagrange multiplier $\lambda_s \geq 0$ should be guaranteed to satisfy the complementarity slackness condition of the power constraint (\ref{eq:sum rate problem Ws b}).
Plugging the obtained transmit beamformers $\mathbf{w}_{s,k_s}^\star, \forall k_s,$ into the power constraint (\ref{eq:sum rate problem Ws b}), the optimal solution of $\lambda_s$ can be easily obtained by one dimensional search methods, e.g., the bisection search method.

\subsubsection{Update $\bm{\Phi}$ and $\mathbf{A}$}
Since the ideal IRS phase-shift $\bm{\Phi}$ is closely associated with the service selection $\mathbf{A}$ in designing the practical IRS reflection $\bm{\Theta}$ for all BSs, a joint design for $\bm{\Phi}$ and $\mathbf{A}$ is necessary to achieve a satisfactory performance, which has not been considered in the existing literature.
Thus, with the fixed variable $\bm{\nu}$, MSE weight $\bm{\mu}$, and transmit beamformer $\mathbf{W}$, the optimization problem for the joint ideal IRS phase-shift $\bm{\Phi}$ and service selection matrix $\mathbf{A}$ design is written by
\begin{subequations}
\label{eq:sum rate phi A}
\begin{align}
    \min \limits_{\bm{\Phi},\mathbf{A}}
    &\sum_{s\in\mathcal{S}}\sum_{k_s\in\mathcal{K}_s}\mu_{s,k_s}\Big[\hspace{-0.1cm}
    \sum_{j\in\mathcal{K}_s}\hspace{-0.1cm}\big|\nu_{s,k_s}^*({\mathbf{h}^{H}_{\text{r},s,k_s}}
    \bm{\Theta}_s\mathbf{G}_{s}\hspace{-0.1cm}+\hspace{-0.1cm}\mathbf{h}_{\text{d},s,k_s}^H)
    \mathbf{w}_{s,j}\big|^2  \nonumber\\
    &~~~~~-2\mathfrak{R}\big\{\nu_{s,k_s}^*({\mathbf{h}^{H}_{\text{r},s,k_s}}
    \bm{\Theta}_s\mathbf{G}_{s} + \mathbf{h}_{\text{d},s,k_s}^H)
    \mathbf{w}_{s,k_s}\big\}\Big]\\
     \textrm{s.t.} &~~(\text{\ref{eq:sum rate problem c}})-(\text{\ref{eq:sum rate problem e}}).
\end{align}
\end{subequations}
For brevity, we define $\mathbf{B}_s$ and $\mathbf{c}_s$, which are presented in (\ref{eq:Bs cs}) on the top of this page, and rearrange problem (\ref{eq:sum rate phi A}) as
\begin{subequations}\label{eq:sum rate phi A reformulate}
\begin{align}\label{eq:sum rate phi A reformulate a}
    \min \limits_{\bm{\Phi},\mathbf{A}}~~&\sum_{s=1}^S\Big( \bm{\theta_s}^H\mathbf{B}_s\bm{\theta_s} - 2\mathfrak{R}\big\{\bm{\theta_s}^H\mathbf{c}_s\big\}\Big)\\
    \textrm{s.t.}~~&\bm{\theta}_s = \exp\left(\jmath\bm{\phi}_s \odot \widetilde{\mathbf{a}}_s\right),\;\; \forall s, \\
   &\phi_{s,m} \in (0, 2\pi], \;\;\forall s,\; m, \\
   \label{eq:sum rate phi A reformulate d}
   &\left\| \mathbf{a}_m \right\|_0 \leq 1, \;\;a_{s,m} \in \{0,1\}, \;\;\forall s,\; m.
\end{align}
\end{subequations}
Due to the joint optimization of the $M$ reflecting elements and the non-convex constraint (\ref{eq:sum rate phi A reformulate d}) of each reflecting element, it is very difficult to directly seek for a optimal solution to problem (\ref{eq:sum rate phi A reformulate}).
Therefore, we attempt to decompose problem (\ref{eq:sum rate phi A reformulate}) into $M$ sub-problems, each of which jointly designs the ideal phase-shifts and service selection of a certain reflecting element with the other reflecting elements fixed.
To this end, we first split the objective function (\ref{eq:sum rate phi A reformulate a}) as
\begin{subequations}\label{eq:unfold theta A}
\begin{align}
   &\sum_{s=1}^S\Big(\bm{\theta_s}^H\mathbf{B}_s\bm{\theta_s} - 2\mathfrak{R}\big\{\bm{\theta_s}^H\mathbf{c}_s\big\}\Big)\\
    &\overset{(b)}{=}\hspace{-0.07cm}\sum_{s=1}^S\hspace{-0.07cm}\sum_{m=1}^M
    \hspace{-0.12cm}\Big[\hspace{-0.07cm}
    \sum_{n=1}^M\mathbf{B}_s(m,n)\theta^*_{s,m}\theta_{s,n}
\hspace{-0.07cm}-\hspace{-0.07cm}2\mathfrak{R}\big\{\theta_{s,m}^*\mathbf{c}_s(m)\big\}\hspace{-0.07cm}\Big]\\
&\overset{(c)}{=}\hspace{-0.05cm}\sum_{s=1}^S \hspace{-0.05cm}\Big[\hspace{-0.05cm}\sum_{n \neq m}\hspace{-0.05cm}\big(\mathbf{B}_s(m,n)\theta_{s,m}^*\theta_{s,n}\hspace{-0.05cm}+ \hspace{-0.05cm}\mathbf{B}_s(n,m)\theta_{s,n}^*\theta_{s,m}\big)\nonumber\\
&~~~~+ \mathbf{B}_s(m,m)|\theta_{s,m}|^2 - 2\mathfrak{R}\big\{\theta_{s,m}^*\mathbf{c}_s(m)\big\}\Big],\\
&\overset{(d)}{=}\hspace{-0.05cm}\sum_{s=1}^S\Big( 2\mathfrak{R}\big\{\zeta_{s,m}\theta_{s,m}^*\big\} + \mathbf{B}_s(m,m)\Big),
\end{align}\end{subequations}
where $\zeta_{s,m} \triangleq \sum_{m \neq n} \mathbf{B}_s(m,n)\theta_{s,n}-\mathbf{c}_s(m)$ for brevity.
In particular, step $(b)$ unfolds the quadratic term of $\bm{\theta}_s$, step $(c)$ separates the reflection coefficients of the $m$-th reflecting elements and the others, and step $(d)$ is derived based on $\mathbf{B}_s = \mathbf{B}_s^H$ and $|\theta_{s,m}| = 1$.

\begin{algorithm}[t]\small
\caption{Joint Transmit Beamforming and Practical IRS Reflection Design for Sum-Rate Maximization Problem}
\label{alg:Algorithm 3}
    \begin{algorithmic}[1]
    \REQUIRE $\mathbf{h}_{\text{r},s,k_s}^H$, $\mathbf{G}_s$, $\mathbf{h}_{\text{d},s,k_s}^H$, $P_s$, $\forall s\in\mathcal{S}$, $\forall k_s\in\mathcal{K}_s$, $\sigma^2$.
    \ENSURE $\mathbf{w}_{s,k_s}^\star, \forall s\in\mathcal{S}, \forall k_s\in\mathcal{K}_s$,\; $\bm{\phi}_m^\star, \forall m\in\mathcal{M}$,\;\;$\mathbf{A}^\star$.
    \STATE {Initialize $\mathbf{A}$, $\bm{\Phi}$, $\mathbf{W}$.}
    \WHILE {no convergence of the objective (\ref{eq:sum rate problem reformualte a})}
    \STATE {Update $\nu_{s,k_s}, \forall s\in\mathcal{S}, \forall k_s\in\mathcal{K}_s$ by (\ref{eq:optimal nu}).}
    \STATE {Update $\mu_{s,k_s}, \forall s\in\mathcal{S}, \forall k_s\in\mathcal{K}_s$ by (\ref{eq:optimal mu}).}
    \STATE {Update $\mathbf{w}_{s,k_s}, \forall s\in\mathcal{S}, \forall k_s\in\mathcal{K}_s$ by (\ref{eq:optimal wsk}).}
    \WHILE {no convergence of the objective (\ref{eq:sum rate phi A reformulate a})}
    \FOR {$m=1:M$}
    \STATE{Obtain $s^\star$ by solving (\ref{eq:sstar problem}).}
    \STATE{Update $\phi_{s^\star,m}$ by (\ref{eq:optimal phi sstarm}).}
    \STATE{Update $\mathbf{a}_m$ by (\ref{eq:update am}).}
    \ENDFOR
    \ENDWHILE
    \ENDWHILE
    \STATE {Return $\mathbf{w}_{s,k_s}^\star, \forall s\in\mathcal{S}, \forall k_s\in\mathcal{K}_s$,\; $\bm{\phi}_m^\star, \forall m\in\mathcal{M}$,\\
    $\mathbf{A}^\star = [\mathbf{a}_1^\star,\ldots,\mathbf{a}_M^\star]$.}
    \end{algorithmic}
\end{algorithm}

Based on the derivations in (\ref{eq:unfold theta A}), when we only consider the $m$-th reflecting element, its corresponding ideal phase-shifts vector $\widetilde{\bm{\phi}}_m \triangleq [\phi_{1,m},\ldots,\phi_{S,m}]^T$ of all BSs and service selection vector $\mathbf{a}_m$ are jointly designed.
Thus, after discarding some constant and irrelevant terms, the $m$-th sub-problem is formulated as
\begin{subequations}\label{eq:phim am}
    \begin{align}\label{eq:phim am a}
    \min \limits_{\widetilde{\bm{\phi}}_m, \mathbf{a}_m}~~&\sum_{s=1}^S \left|\zeta_{s,m}\right|\cos\left(\angle\zeta_{s,m} - \phi_{s,m}a_{s,m}\right)\\
     \textrm{s.t.}~~&\phi_{s,m} \in (0, 2\pi], \;\;\forall s, \\
     \label{eq:phim am c}
     &\left\| \mathbf{a}_m \right\|_0 \leq 1, \;\;a_{s,m} \in \{0,1\}, \;\;\forall s.
   \end{align}
\end{subequations}
We observe that the binary vector $\mathbf{a}_m$ contains at most one non-zero entry according to constraint (\ref{eq:phim am c}).
If $a_{s,m}=0$, there is no need to optimize $\phi_{s,m}$ since it cannot change the objective value.
Meanwhile, if $a_{s,m}=1$, there must exist a $\phi_{s,m} = \pi+\angle\zeta_{s,m}$ that achieves the minimum value $-|\zeta_{s,m}|$ of the $s$-th term in (\ref{eq:phim am a}).
Therefore, the optimal solution $\mathbf{a}_m^\star$ to problem (\ref{eq:phim am}) is a non-zero vector.
We assume the $s^\star$-th entry of $\mathbf{a}_m^\star$ is 1 and thus only the $s^\star$-th entry of $\widetilde{\bm{\phi}}_m$ is updated by
\begin{equation}\label{eq:optimal phi sstarm}
\phi_{s^\star,m} = \pi + \angle\zeta_{s^\star,m}.
\end{equation}
Then, the optimization problem (\ref{eq:phim am}) is converted into
\begin{subequations}\label{eq:sstar problem}
\begin{align}
\min \limits_{s^\star}~~& \sum_{\substack{s=1\\s\neq s^\star}}^S \left|\zeta_{s,m}\right|\cos\angle\zeta_{s,m}
-\left|\zeta_{s^\star,m}\right|\\
\label{eq:s feasible set}
\text{s.t.}~~&s^\star = 1,2,\ldots,S,
\end{align}
\end{subequations}
which can be easily solved by exhaustively searching its feasible set (\ref{eq:s feasible set}).
After finding $s^\star$, the optimal ideal phase-shift $\phi_{s^\star,m}$ is updated by (\ref{eq:optimal phi sstarm}) and the optimal service selection vector $\mathbf{a}_m^\star$ is determined by
\begin{equation}\label{eq:update am}
a_{s^\star,m} = 1, \;\; a_{s,m} = 0,\;\forall s\in\mathcal{S}, s\neq s^\star.
\end{equation}

\begin{figure}[t]
\centering
  \includegraphics[height = 1.5 in]{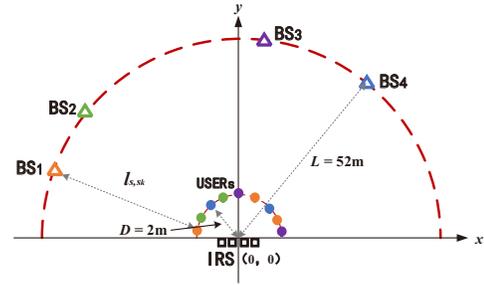}
  \caption{An illustration of the relative position among the BSs, IRS, and users.}\label{fig:model}
  \vspace{-0.5 cm}
\end{figure}

\subsection{Summary and Complexity Analysis}
\textit{Summary}:
Based on the above derivations, the joint transmit beamforming and practical IRS reflection design for sum-rate maximization problem is straightforward and summarized in Algorithm 2.
With an appropriate initialization, the variable $\bm{\nu}$, the MSE weight $\bm{\mu}$, the transmit beamformer $\mathbf{W}$, and the practical IRS reflection $\bm{\Theta} = \exp(\bm{\Phi}\odot\mathbf{A}),$ are iteratively updated until the convergence is met.
Since the objective function value at each step is nondecreasing, we can guarantee that the BCD method can strictly converge to a local optimum point.
In addition, in the initialization, the service selection matrix $\mathbf{A}$ and the ideal phase-shift $\bm{\Phi}$ are randomly generated from their corresponding feasible sets, and the transmit beamformer of each BS adopts the typical MMSE beamformer.

\textit{Complexity analysis}: In each iteration, updating $\bm{\nu}$ has a complexity of $\mathcal{O}\{\sum_{s \in \mathcal{S}}K_s(K_s+1)N_\text{t}M^2\}$; updating $\bm{\mu}$ has a complexity of $\mathcal{O}\{\sum_{s \in \mathcal{S}}K_s^2N_\text{t}M^2\}$; updating transmit beamformer $\mathbf{W}$ requires about $\mathcal{O}\{\sum_{s \in \mathcal{S}}N_\text{t}K_s(3M^2+N^2_\text{t})\}$ operations.
Finally, the order of complexity for updating ideal IRS phase-shifts $\mathbf{\Phi}$ and service selection matrix $\mathbf{A}$ is about
$\mathcal{O}\{\sum_{s \in \mathcal{S}}[(M-1)(K_s^3M^2N_\mathrm{t}^2 + K_s^3MN_\mathrm{t}^2)]\}$,
Therefore, the total complexity of the proposed algorithm is $
\mathcal{O}\big\{\sum_{s \in \mathcal{S}}K_s
\big[(2K_s+1)N_\text{t}M^2+K_s(3M^2+N_\text{t}^2)+K_s^2(MN_\text{t}^2(M-1)(M+N_\text{t}))\big]\big\}$.

\section{Simulation Results}
In this section, extensive simulation results are presented to demonstrate the significance of using the proposed practical IRS reflection model in the IRS-assisted multi-cell multi-band system and the effectiveness of our proposed algorithms.
All simulations are implemented on a computer with Intel(R) Core(TM) i7-9700 CPU, NVIDIA GeForce GT-710 GPU, Matlab Version 2020b, CVX toolbox Version 2.2, and Manopt toolbox Version 7.0.
We assume that the multi-cell multi-band system consists of $S=3$ or $S=4$ cells/BSs.
In each cell, the BS equipped with $N_\text{t} = 16$ antennas serves $K = K_s = 3, \forall s,$ single-antenna users.
An IRS composed of $M = 64$ reflecting elements is deployed to assist the downlink communications for the considered system.
The noise power is set as $\sigma^2 = -70$dBm.
The QoS requirement of each user and the power budget of each BS are the same, i.e., $\Gamma = \Gamma_{s,k_s} = 5\textrm{dB}, \forall s, \forall k_s$, and $P = P_s = -5\textrm{dB}, \forall s$.
In addition, the distance-dependent channel path loss\footnote{This assumption is widely used in the existing works [7], [14], [15], [20], [22]-[24], [26]-[30]. Moreover, proposed practical IRS reflection model and associated algorithms are also suitable for other realistic propagation environments in \cite{3GPP}.} is modeled as $\eta(d) = C_0(\frac{d}{d_0})^{-\alpha}$, where $C_0 = -30$dB denotes the signal attenuation at the reference distance $d_0 = 1$m, and $\alpha$ denotes the path loss exponent.
We set the path loss exponents for the BS-IRS, IRS-user, and BS-user channels as 2.5, 2.8, and 3.5, respectively.

\begin{figure}[t]
\centering
  \vspace{0.25 cm}
  \includegraphics[width = 3.5 in]{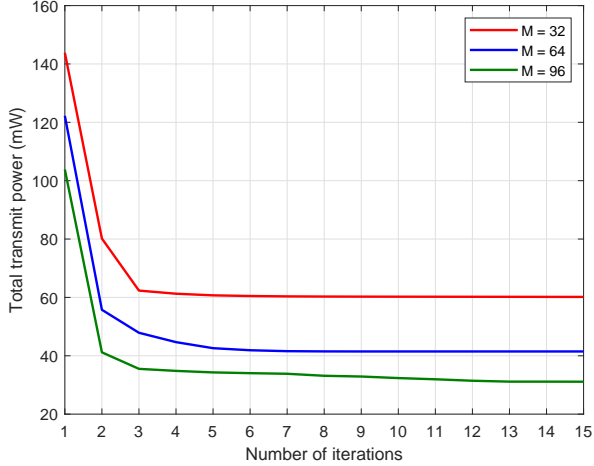}\\
  \vspace{0.25 cm}
  \caption{Transmit power versus the number of iterations ($N_\text{t} = 16$, $S = 3$, $K = 3$, $\Gamma = 5$dB).}\label{fig:power_iter}
  \vspace{-0.4 cm}
\end{figure}
\begin{figure}[t]
  \hspace{0.1 cm}
  \includegraphics[width = 3.7 in]{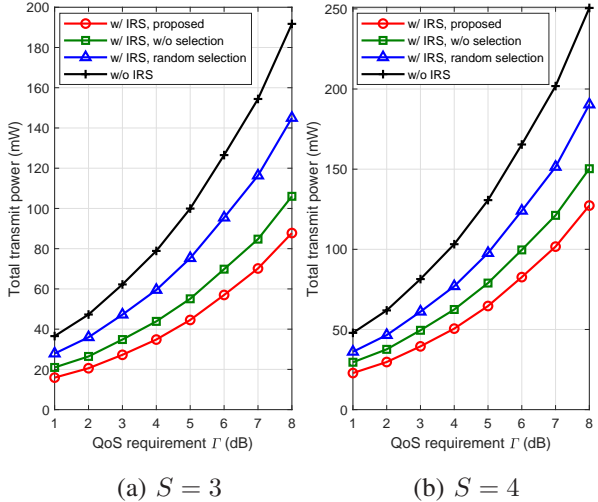}\\
  \vspace{0.1 cm}
  \hspace{1.5 cm} (a) $S = 3$ \hspace{2.3 cm} (b) $S = 4$
  \caption{Total transmit power versus the QoS requirement $\Gamma$ ($M = 64$, $N_\text{t} = 16$, $K = 3$).}\label{fig:power_snr}
  \vspace{-0.4 cm}
\end{figure}
\begin{figure}[t]
  \hspace{0.1 cm}
  \includegraphics[width = 3.7 in]{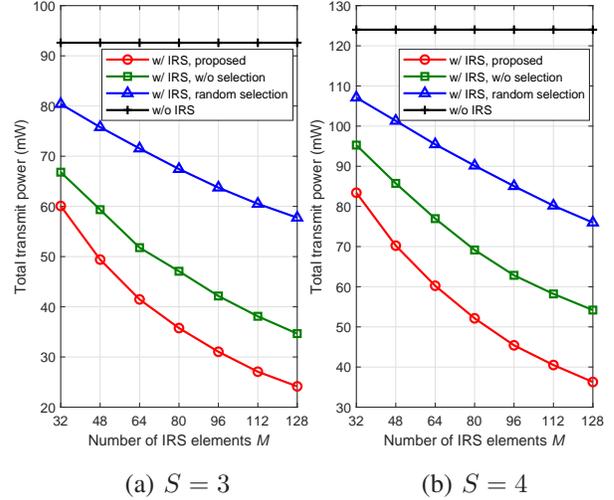}\\
  \vspace{0.1 cm}
  \hspace{1.5 cm} (a) $S = 3$ \hspace{2.3 cm} (b) $S = 4$
  \caption{Total transmit power versus the number of IRS reflecting elements $M$ ($\Gamma = 5$dB, $N_\text{t} = 16$, $K = 3$).}\label{fig:power_m}
  \vspace{-0.4 cm}
\end{figure}
\begin{figure}[t]
  \hspace{0.1 cm}
  \includegraphics[width = 3.7 in]{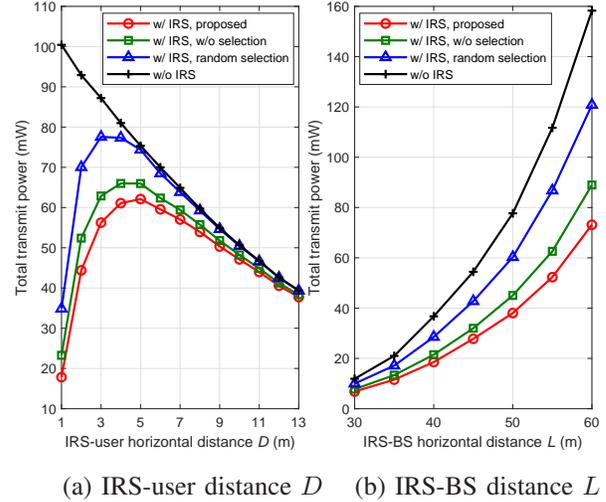}\\
  \vspace{0.1 cm}
  \hspace{0.9 cm}(a) IRS-user distance $D$
  \hspace{0.3 cm}(b) IRS-BS distance $L$
  \vspace{0.1 cm}
  \caption{Total transmit power versus the distance between BSs, IRS, and users ($M = 64$, $\Gamma = 5$dB, $N_\text{t} = 16$, $S = 3$, $K = 3$).}\label{fig:power_d}
  \vspace{-0.4 cm}
\end{figure}

A two-dimensional coordinate system is shown in Fig. \ref{fig:model} to demonstrate the position relationship of different devices in the considered systems from a top-down view.
For both cases of $S=3$ BSs and $S=4$ BSs, the IRS is located at $(0,0)$, and the BSs are randomly distributed at a distance of $L = 52$m away from the IRS.
Since the IRS is deployed to enhance the transmission quality for the edge users, we assume that the users in each cell are randomly distributed $D = 2$m away from the IRS.

\begin{figure}[t]
\centering
  \vspace{0.25 cm}
  \includegraphics[width = 3.5 in]{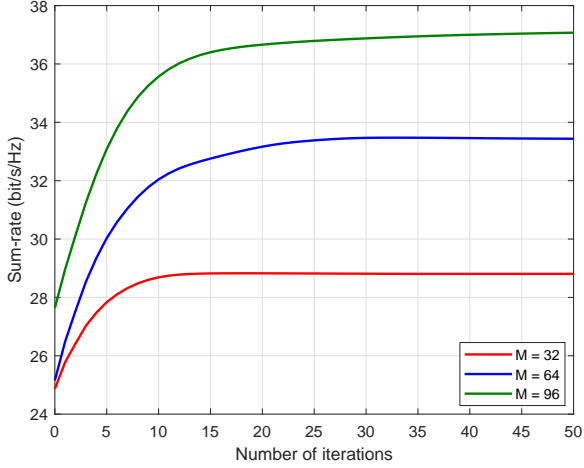}\\
  \vspace{0.25 cm}
  \caption{Sum-rate versus the number of iterations
  ($P = -5$dB, $N_\text{t} = 16$, $S = 3$, $K = 3$).}\label{fig:rate_iter}
  \vspace{-0.4 cm}
\end{figure}
\begin{figure}[t]
\hspace{0.1 cm}
  \includegraphics[width = 3.7 in]{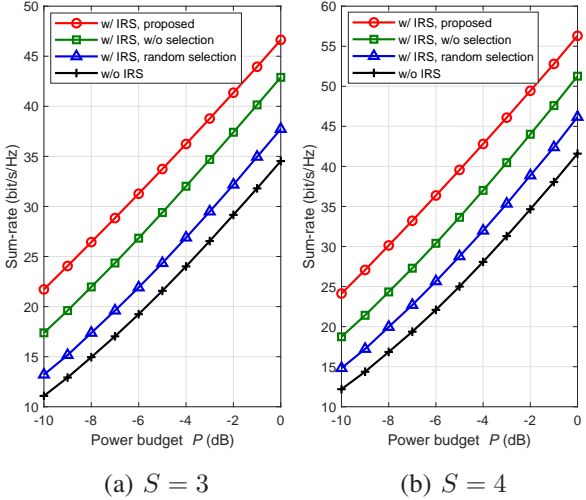}\\
  \vspace{0.1 cm}
  \hspace{1.5 cm} (a) $S = 3$ \hspace{2.3 cm} (b) $S = 4$
  \caption{Sum-rate versus the power budget $P$
  ($M = 64$, $N_\text{t} = 16$, $K = 3$).}\label{fig:rate_snr}
  \vspace{-0.4 cm}
\end{figure}
\begin{figure}[t]
\hspace{0.1 cm}
  \includegraphics[width = 3.7 in]{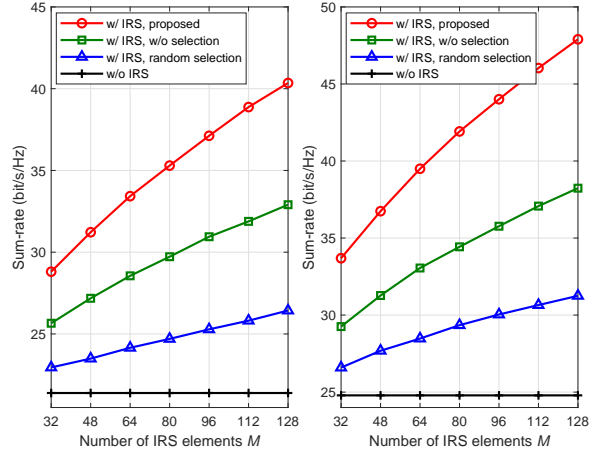}\\
  \vspace{0.1 cm}
  \hspace{1.5 cm} (a) $S = 3$ \hspace{2.3 cm} (b) $S = 4$
  \caption{Sum-rate versus the number of IRS reflecting elements $M$  ($P = -5$dB, $N_\text{t} = 16$, $K = 3$).}\label{fig:rate_M}
  \vspace{-0.4 cm}
\end{figure}
\begin{figure}[t]
\hspace{0.1 cm}
  \includegraphics[width = 3.7 in]{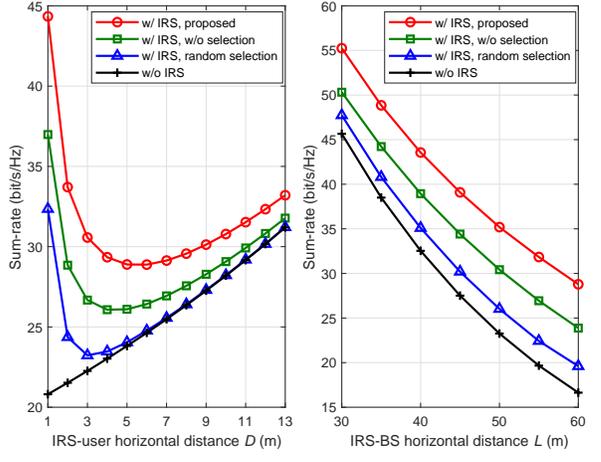}\\
  \vspace{0.1 cm}
  \hspace{0.9 cm}(a) IRS-user distance $D$
  \hspace{0.3 cm}(b) IRS-BS distance $L$
  \vspace{0.1 cm}
  \caption{Sum-rate versus the distance between BSs, IRS, and users ($M = 64$, $P = -5$dB, $N_\text{t} = 16$,
  $S = 3$, $K = 3$).}\label{fig:rate_d}
  \vspace{-0.4 cm}
\end{figure}
\subsection{Power Minimization Problem}
In this subsection, we show the simulation results for the power minimization problem in Figs. \ref{fig:power_iter}-\ref{fig:power_d}.
The transmit power versus the number of iterations is first presented in Fig. \ref{fig:power_iter} to show the convergence of the proposed algorithm.
It can be observed that the convergence can be met within 12 iterations under different settings and the scheme with lesser IRS reflecting elements has faster convergence due to the reduced dimension of variables.
These convergence results support the low-complexity implementation.

Fig. \ref{fig:power_snr} shows the total transmit power versus the QoS requirement $\Gamma$ for the cases that $S = 3$ and $S = 4$.
Our algorithm based on the proposed practical IRS reflection model in this paper is denoted as ``w/ IRS, proposed''.
The practical IRS reflection model for the considered multi-cell multi-band systems exhibits that each reflecting element is independently selected to serve a certain BS.
Therefore, for comparison we also include \textit{i)} the scenario that all reflecting elements are selected to serve the same BS, i.e., the IRS provides tunable phase-shifts for a certain BS while exhibits fixed $0^\circ$ phase-shifts for other BSs, which is denoted as ``w/ IRS, w/o selection''; \textit{ii)} the scheme that each reflecting element is randomly selected to serve a certain BS, which is denoted as ``w/ IRS, random selection''; \textit{iii)} the case without deploying the IRS, which is denoted as ``w/o IRS''.
We can easily observe from Fig. \ref{fig:power_snr} that the proposed algorithm requires less transmit power than the ``w/ IRS, random selection'' and ``w/o IRS'' schemes for all transmit power ranges, which validates the advantages of deploying IRS in wireless communication systems.
Moreover, the proposed algorithm outperforms the ``w/ IRS, w/o selection'' scheme, which verifies the importance of joint designing the transmit beamforming, ideal phase-shift design, and IRS service selection.

Then, the transmit power versus the number of IRS reflecting elements $M$ is plotted in Fig. \ref{fig:power_m}.
A similar conclusion can be drawn as that from Fig. \ref{fig:power_snr}.
We also observe that the transmit power of all schemes decreases with the increasing number of IRS reflecting elements, and the power reduction of our proposed algorithm is more remarkable, which validates the advancement of the proposed practical model and design algorithm in playing the role of the IRS.

To show the impact of IRS location, Fig. \ref{fig:power_d} illustrates the total transmit power as functions of the IRS-user distance $D$ and IRS-BS distance $L$.
It can be observed in Fig. \ref{fig:power_d} (a) that for the schemes with IRS, when the users move away from the IRS towards the BSs, the transmit power increases at first, because the reflected signals from the IRS becomes weaker, and then decreases thanks to stronger signals from BSs.
It is worth noting that our proposed algorithm always has better performance and service coverage.
For the ``w/o IRS'' scheme, the transmit power keeps decreasing since the users move closer to their corresponding BSs.
When users are randomly distributed at a distance of
$D =2$m away from the IRS, the total transmit power versus the IRS-BS distance $L$ is shown in Fig. \ref{fig:power_d} (b).
The longer distance between BS and IRS, the more significant performance improvement of deploying IRS.
Moreover, our proposed algorithm always outperforms the others under different IRS-user and IRS-BS distance settings.

\begin{figure}[t]
  \hspace{0.1 cm}
  \includegraphics[width = 3.7 in]{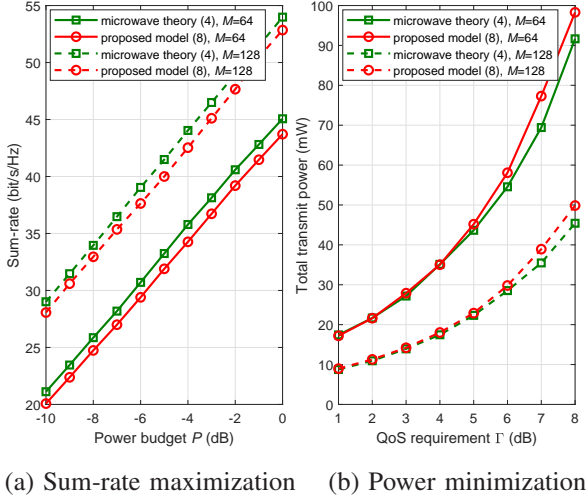}\\
  \vspace{0.1 cm}
  \hspace{0.3 cm}(a) Sum-rate maximization \hspace{0.2 cm} (b) Power minimization
  \vspace{0.1 cm}
  \caption{Performance comparison between the practical response and the proposed model  ($M = 64$, $N_\text{t} = 16$, $S = 3$, $K = 3$).}
  \vspace{-0.5 cm}
\end{figure}
\subsection{Sum-rate Maximization Problem}
In this subsection, the simulation results for the sum-rate maximization problem are demonstrated in Figs. \ref{fig:rate_iter}-\ref{fig:rate_d}.
A similar fast convergence performance as that for the power minimization problem is also observed in Fig. \ref{fig:rate_iter}.

Fig. \ref{fig:rate_snr} shows the sum-rate versus the power budget $P$ for the scenarios that $S = 3$ and $S = 4$.
It can be seen that more transmit power provides a larger sum-rate for all schemes and the proposed algorithm always significantly outperforms the others under different settings.
Specifically, our proposed algorithm can provide about 3dB gain compared with the scheme without BS selection, and more than 4dB gain compared with the scheme with random BS selection.
The sum-rate versus the number of IRS reflecting elements $M$ is illustrated in Fig. \ref{fig:rate_M}.
Similar conclusion can be drawn as that from Fig. \ref{fig:power_m}.
Note that our proposed algorithm has a larger performance improvement with the increase of $M$. This is because more reflecting elements can provide larger passive beamforming gain, while the practical model and proposed design algorithm can utilize all IRS elements more efficiently.
The sum-rate versus the IRS-user distance $D$ and IRS-BS distance $L$ are presented in Fig. \ref{fig:rate_d} (a) and (b), respectively. The impact of the IRS locations on the system performance is consistent with that in Fig. 8, which further demonstrates the importance of IRS deployment with the proposed design algorithm in practical systems.

\subsection{Analysis of Model Errors}
In order to illustrate the impact of the phase approximation of the proposed IRS model in Table. I on the degradation of system performance, Fig. 13 compares the performances using the microwave theory based actual response expressed in (4), and the proposed simplified model (8).
Specifically, when the microwave theory (4) is used, the phase-shift response of other frequencies is not approximated as unchanged, but also varied accordingly. However, the microwave theory (4) is extremely complicated and no existing algorithm can be directly adopted.
Thus, we iteratively design the IRS phase-shift for each element by one-dimensional exhaustive search with fixed other elements. As seen from Fig. 13, the microwave theory based actual response (4) can provide slightly better performance than our proposed simplified model.
However, the exhaustive search algorithm that enables the design under complex nonlinear equation (4) causes significantly higher computational complexity.
For example, for a 64-element IRS-assisted system, the beamforming design with the microwave theory (4) takes an average of 46.217 seconds using the exhaustive search method, while the proposed simplified model (8) only requires an average of 1.072 seconds using our proposed beamforming design algorithm.
Therefore, we can conclude that our proposed model can facilitate the beamforming design with negligible performance loss.

\section{Conclusions}
In this paper, we first derived a practical yet tractable IRS reflection model for an IRS-assisted multi-cell multi-band system.
Based on the proposed model, we investigated the joint transmit beamforming and IRS reflection designs for both power minimization and sum-rate maximization problems.
Efficient algorithms were proposed to solve them by exploiting SOCP, Riemannian manifold, WMMSE, BCD, and the proposed efficient search algorithms
Simulation results demonstrated significant performance improvement of the proposed algorithms, which confirmed the importance of using the proposed practical reflection model and the joint beamforming and IRS reflection designs in IRS-assisted multi-cell multi-band systems.
Moreover, there are many issues of IRS-assisted systems with the practical reflection model worth being investigated in future works, including more sophisticated beamforming design algorithms, fast channel estimation, influence of CSI errors, as well as learning-based methods, etc.

\end{document}